
%
%
%
%
%
%
%
\nonstopmode
\catcode`\@=11 
%
%
%
\font\seventeenrm=cmr17

\font\twelverm=cmr12
\font\ninerm=cmr9
\font\sixrm=cmr6

\font\seventeenbf=cmbx12 at 17pt
\font\fourteenbf=cmbx12 at 14pt
\font\twelvebf=cmbx12
\font\ninebf=cmbx9
\font\sixbf=cmbx6

\font\seventeeni=cmmi12 at 17pt             \skewchar\seventeeni='177
\font\fourteeni=cmmi12 at 14pt              \skewchar\fourteeni='177
\font\twelvei=cmmi12                        \skewchar\twelvei='177
\font\ninei=cmmi9                           \skewchar\ninei='177
\font\sixi=cmmi6                            \skewchar\sixi='177

\font\seventeensy=cmsy10 scaled\magstep3    \skewchar\seventeensy='60
\font\fourteensy=cmsy10 scaled\magstep2     \skewchar\fourteensy='60
\font\twelvesy=cmsy10 at 12pt               \skewchar\twelvesy='60
\font\ninesy=cmsy9                          \skewchar\ninesy='60
\font\sixsy=cmsy6                           \skewchar\sixsy='60

\font\seventeenex=cmex10 scaled\magstep3
\font\fourteenex=cmex10 scaled\magstep2
\font\twelveex=cmex10 at 12pt

\font\ninex=cmex10 at 9pt
\font\sevenex=cmex10 at 9pt
\font\sixex=cmex10 at 6pt
\font\fivex=cmex10 at 5pt

\font\seventeensl=cmsl10 scaled\magstep3
\font\fourteensl=cmsl10 scaled\magstep2
\font\twelvesl=cmsl10 scaled\magstep1
\font\ninesl=cmsl10 at 9pt
\font\sevensl=cmsl10 at 7pt
\font\sixsl=cmsl10 at 6pt
\font\fivesl=cmsl10 at 5pt

\font\seventeenit=cmti12 scaled\magstep2
\font\fourteenit=cmti12 scaled\magstep1
\font\twelveit=cmti12

\font\seventeentt=cmtt12 scaled\magstep2
\font\fourteentt=cmtt12 scaled\magstep1
\font\twelvett=cmtt12

\font\seventeencp=cmcsc10 scaled\magstep3
\font\fourteencp=cmcsc10 scaled\magstep2
\font\twelvecp=cmcsc10 scaled\magstep1
\font\tencp=cmcsc10

\newfam\cpfam

\font\seventeenss=cmss17
\font\fourteenss=cmss12 at 14pt
\font\twelvess=cmss12
\font\tenss=cmss10
\font\niness=cmss9

\font\sevenss=cmss8 at 7pt
\font\sixss=cmss8 at 6pt
\font\fivess=cmss8 at 5pt
\newfam\ssfam
\newdimen\b@gheight             \b@gheight=12pt
\newcount\f@ntkey               \f@ntkey=0
\def\f@m{\afterassignment\samef@nt\f@ntkey=}
\def\samef@nt{\fam=\f@ntkey \the\textfont\f@ntkey\relax}
\def\rm{\f@m0 }
\def\mit{\f@m1 }         
\def\cal{\f@m2 }
\def\it{\f@m\itfam}
\def\sl{\f@m\slfam}
\def\bf{\f@m\bffam}
\def\tt{\f@m\ttfam}
\def\ssf{\f@m\ssfam}
\def\caps{\f@m\cpfam}
\def\seventeenpoint{\relax
    \textfont0=\seventeenrm          \scriptfont0=\twelverm
      \scriptscriptfont0=\ninerm
    \textfont1=\seventeeni           \scriptfont1=\twelvei
      \scriptscriptfont1=\ninei
    \textfont2=\seventeensy          \scriptfont2=\twelvesy
      \scriptscriptfont2=\ninesy
    \textfont3=\seventeenex          \scriptfont3=\twelveex
      \scriptscriptfont3=\ninex
    \textfont\itfam=\seventeenit    
    \textfont\slfam=\seventeensl    
      \scriptscriptfont\slfam=\ninesl
    \textfont\bffam=\seventeenbf     \scriptfont\bffam=\twelvebf
      \scriptscriptfont\bffam=\ninebf
    \textfont\ttfam=\seventeentt
    \textfont\cpfam=\seventeencp
    \textfont\ssfam=\seventeenss     \scriptfont\ssfam=\twelvess
      \scriptscriptfont\ssfam=\niness
    \samef@nt
    \b@gheight=17pt
    \setbox\strutbox=\hbox{\vrule height 0.85\b@gheight
                                depth 0.35\b@gheight width\z@ }}
\def\fourteenpoint{\relax
    \textfont0=\fourteencp          \scriptfont0=\tenrm
      \scriptscriptfont0=\sevenrm
    \textfont1=\fourteeni           \scriptfont1=\teni
      \scriptscriptfont1=\seveni
    \textfont2=\fourteensy          \scriptfont2=\tensy
      \scriptscriptfont2=\sevensy
    \textfont3=\fourteenex          \scriptfont3=\twelveex
      \scriptscriptfont3=\tenex
    \textfont\itfam=\fourteenit     \scriptfont\itfam=\tenit
    \textfont\slfam=\fourteensl     \scriptfont\slfam=\tensl
      \scriptscriptfont\slfam=\sevensl
    \textfont\bffam=\fourteenbf     \scriptfont\bffam=\tenbf
      \scriptscriptfont\bffam=\sevenbf
    \textfont\ttfam=\fourteentt
    \textfont\cpfam=\fourteencp
    \textfont\ssfam=\fourteenss     \scriptfont\ssfam=\tenss
      \scriptscriptfont\ssfam=\sevenss
    \samef@nt
    \b@gheight=14pt
    \setbox\strutbox=\hbox{\vrule height 0.85\b@gheight
                                depth 0.35\b@gheight width\z@ }}
\def\twelvepoint{\relax
    \textfont0=\twelverm          \scriptfont0=\ninerm
      \scriptscriptfont0=\sixrm
    \textfont1=\twelvei           \scriptfont1=\ninei
      \scriptscriptfont1=\sixi
    \textfont2=\twelvesy           \scriptfont2=\ninesy
      \scriptscriptfont2=\sixsy
    \textfont3=\twelveex          \scriptfont3=\ninex
      \scriptscriptfont3=\sixex
    \textfont\itfam=\twelveit    
    \textfont\slfam=\twelvesl    
      \scriptscriptfont\slfam=\sixsl
    \textfont\bffam=\twelvebf     \scriptfont\bffam=\ninebf
      \scriptscriptfont\bffam=\sixbf
    \textfont\ttfam=\twelvett
    \textfont\cpfam=\twelvecp
    \textfont\ssfam=\twelvess     \scriptfont\ssfam=\niness
      \scriptscriptfont\ssfam=\sixss
    \samef@nt
    \b@gheight=12pt
    \setbox\strutbox=\hbox{\vrule height 0.85\b@gheight
                                depth 0.35\b@gheight width\z@ }}
\def\tenpoint{\relax
    \textfont0=\tenrm          \scriptfont0=\sevenrm
      \scriptscriptfont0=\fiverm
    \textfont1=\teni           \scriptfont1=\seveni
      \scriptscriptfont1=\fivei
    \textfont2=\tensy          \scriptfont2=\sevensy
      \scriptscriptfont2=\fivesy
    \textfont3=\tenex          \scriptfont3=\sevenex
      \scriptscriptfont3=\fivex
    \textfont\itfam=\tenit     \scriptfont\itfam=\seveni
    \textfont\slfam=\tensl     \scriptfont\slfam=\sevensl
      \scriptscriptfont\slfam=\fivesl
    \textfont\bffam=\tenbf     \scriptfont\bffam=\sevenbf
      \scriptscriptfont\bffam=\fivebf
    \textfont\ttfam=\tentt
    \textfont\cpfam=\tencp
    \textfont\ssfam=\tenss     \scriptfont\ssfam=\sevenss
      \scriptscriptfont\ssfam=\fivess
    \samef@nt
    \b@gheight=10pt
    \setbox\strutbox=\hbox{\vrule height 0.85\b@gheight
                                depth 0.35\b@gheight width\z@ }}
%
%
%
\normalbaselineskip = 15pt plus 0.2pt minus 0.1pt 
\normallineskip = 1.5pt plus 0.1pt minus 0.1pt
\normallineskiplimit = 1.5pt
\newskip\normaldisplayskip
\normaldisplayskip = 15pt plus 5pt minus 10pt 
\newskip\normaldispshortskip
\normaldispshortskip = 6pt plus 5pt
\newskip\normalparskip
\normalparskip = 6pt plus 2pt minus 1pt
\newskip\skipregister
\skipregister = 5pt plus 2pt minus 1.5pt
\newif\ifsingl@    \newif\ifdoubl@
\newif\iftwelv@    \twelv@true
\def\singlespace{\singl@true\doubl@false\spaces@t}
\def\doublespace{\singl@false\doubl@true\spaces@t}
\def\normalspace{\singl@false\doubl@false\spaces@t}
\def\Tenpoint{\tenpoint\twelv@false\spaces@t}
\def\Twelvepoint{\twelvepoint\twelv@true\spaces@t}
\def\spaces@t{\relax
      \iftwelv@ \ifsingl@\subspaces@t3:4;\else\subspaces@t1:1;\fi
       \else \ifsingl@\subspaces@t3:5;\else\subspaces@t4:5;\fi \fi
      \ifdoubl@ \multiply\baselineskip by 5
         \divide\baselineskip by 4 \fi }
\def\subspaces@t#1:#2;{
      \baselineskip = \normalbaselineskip
      \multiply\baselineskip by #1 \divide\baselineskip by #2
      \lineskip = \normallineskip
      \multiply\lineskip by #1 \divide\lineskip by #2
      \lineskiplimit = \normallineskiplimit
      \multiply\lineskiplimit by #1 \divide\lineskiplimit by #2
      \parskip = \normalparskip
      \multiply\parskip by #1 \divide\parskip by #2
      \abovedisplayskip = \normaldisplayskip
      \multiply\abovedisplayskip by #1 \divide\abovedisplayskip by #2
      \belowdisplayskip = \abovedisplayskip
      \abovedisplayshortskip = \normaldispshortskip
      \multiply\abovedisplayshortskip by #1
        \divide\abovedisplayshortskip by #2
      \belowdisplayshortskip = \abovedisplayshortskip
      \advance\belowdisplayshortskip by \belowdisplayskip
      \divide\belowdisplayshortskip by 2
      \smallskipamount = \skipregister
      \multiply\smallskipamount by #1 \divide\smallskipamount by #2
      \medskipamount = \smallskipamount \multiply\medskipamount by 2
      \bigskipamount = \smallskipamount \multiply\bigskipamount by 4 }
\def\normalbaselines{ \baselineskip=\normalbaselineskip
   \lineskip=\normallineskip \lineskiplimit=\normallineskip
   \iftwelv@\else \multiply\baselineskip by 4 \divide\baselineskip by 5
     \multiply\lineskiplimit by 4 \divide\lineskiplimit by 5
     \multiply\lineskip by 4 \divide\lineskip by 5 \fi }
\Twelvepoint  
%
\interlinepenalty=50
\interfootnotelinepenalty=5000
\predisplaypenalty=9000
\postdisplaypenalty=500
\hfuzz=1pt
\vfuzz=0.2pt
\dimen\footins=24 truecm 
\voffset=5 truemm 
%
%
%
%
%
%
\def\footnote#1{\edef\@sf{\spacefactor\the\spacefactor}#1\@sf
      \insert\footins\bgroup\singl@true\doubl@false\Tenpoint
      \interlinepenalty=\interfootnotelinepenalty \let\par=\endgraf
        \leftskip=\z@skip \rightskip=\z@skip
        \splittopskip=10pt plus 1pt minus 1pt \floatingpenalty=20000
        \smallskip\item{#1}\bgroup\strut\aftergroup\@foot\let\next}
\skip\footins=\bigskipamount 
\dimen\footins=24truecm 
\newcount\fnotenumber
\def\clearfnotenumber{\fnotenumber=0}
\def\fnote{\advance\fnotenumber by1 \footnote{$^{\the\fnotenumber}$}}
\clearfnotenumber
%
%
\newcount\secnumber
\newcount\appnumber
\newif\ifs@c 
\newif\ifs@cd 
\s@cdtrue 
\def\unsectioned{\s@cdfalse}
\def\clearappnumber{\appnumber=64}
\def\clearsecnumber{\secnumber=0}
\newskip\sectionskip         \sectionskip=\medskipamount
\newskip\headskip            \headskip=8pt plus 3pt minus 3pt
\newdimen\sectionminspace    \sectionminspace=10pc
\newdimen\referenceminspace  \referenceminspace=25pc
\def\Titlestyle#1{\par\begingroup \interlinepenalty=9999
     \leftskip=0.02\hsize plus 0.23\hsize minus 0.02\hsize
     \rightskip=\leftskip \parfillskip=0pt
     \advance\baselineskip by 0.5\baselineskip
     \hyphenpenalty=9000 \exhyphenpenalty=9000
     \tolerance=9999 \pretolerance=9000
     \spaceskip=0.333em \xspaceskip=0.5em
     \iftwelv@\seventeenpoint\else\fourteenpoint\fi
   \noindent #1\par\endgroup }
\def\titlestyle#1{\par\begingroup \interlinepenalty=9999
     \leftskip=0.02\hsize plus 0.23\hsize minus 0.02\hsize
     \rightskip=\leftskip \parfillskip=0pt
     \hyphenpenalty=9000 \exhyphenpenalty=9000
     \tolerance=9999 \pretolerance=9000
     \spaceskip=0.333em \xspaceskip=0.5em
     \iftwelv@\fourteenpoint\else\twelvepoint\fi
   \noindent #1\par\endgroup }
%
\def\spacecheck#1{\dimen@=\pagegoal\advance\dimen@ by -\pagetotal
   \ifdim\dimen@<#1 \ifdim\dimen@>0pt \vfil\break \fi\fi}
\def\section#1{\cleareqnumber \s@ctrue \global\advance\secnumber by1
   \message{Section \the\secnumber: #1}
   \par \ifnum\the\lastpenalty=30000\else
   \penalty-200\vskip\sectionskip \spacecheck\sectionminspace\fi
   \noindent {\caps\enspace\S\the\secnumber\quad #1}\par
   \nobreak\vskip\headskip \penalty 30000 }
\def\subsection#1{\par
   \ifnum\the\lastpenalty=30000\else \penalty-100\smallskip
   \spacecheck\sectionminspace\fi
   \noindent\undertext{#1}\enspace \vadjust{\penalty5000}}

\def\undertext#1{\vtop{\hbox{#1}\kern 1pt \hrule}}
\def\subsubsection#1{\par
   \ifnum\the\lastpenalty=30000\else \penalty-100\smallskip \fi
   \noindent\hbox{#1}\enspace \vadjust{\penalty5000}}

\def\appendix#1{\cleareqnumber \s@cfalse \global\advance\appnumber by1
   \message{Appendix \char\the\appnumber: #1}
   \par \ifnum\the\lastpenalty=30000\else
   \penalty-200\vskip\sectionskip \spacecheck\sectionminspace\fi
   \noindent {\caps\enspace Appendix \char\the\appnumber\quad #1}\par
   \nobreak\vskip\headskip \penalty 30000 }
\clearsecnumber
\clearappnumber
%
%
\def\ack{\par\penalty-100\medskip \spacecheck\sectionminspace
   \line{\iftwelv@\fourteencp\else\twelvecp\fi\hfil ACKNOWLEDGEMENTS\hfil}%
\nobreak\vskip\headskip }
\def\refs{\begingroup \par\penalty-100\medskip \spacecheck\sectionminspace
   \line{\iftwelv@\fourteencp\else\twelvecp\fi\hfil REFERENCES\hfil}%
\nobreak\vskip\headskip \frenchspacing }
\def\endrefs{\par\endgroup}
%
\newcount\refnumber
\def\clearrefnumber{\refnumber=0}  \clearrefnumber
\newtoks\temptok
\newwrite\R@fs                              
\immediate\openout\R@fs=\jobname.references 
\def\closerefs{\immediate\closeout\R@fs} 
\def\refsout{\closerefs\refs\input\jobname.references\endrefs}
\def\refitem#1{\item{{\bf #1}}}
\def\ifundefined#1{\expandafter\ifx\csname#1\endcsname\relax}
\def\gobble#1{}
\def\Ref#1{\ifundefined{\expandafter\gobble\string #1}%
\global\advance\refnumber by1%
\xdef#1{[\the\refnumber]}%
\temptok={\csname\expandafter\gobble\string#1r\endcsname}
\immediate\write\R@fs{\noexpand\refitem{#1} \the\temptok}\fi%
{\bf #1}}
%
%
\newcount\eqnumber
\def\cleareqnumber{\eqnumber=0}
\newif\ifal@gn \al@gnfalse  
\def\veqnalign#1{\al@gntrue \vbox{\eqalignno{#1}} \al@gnfalse}
\def\eqnalign#1{\al@gntrue \eqalignno{#1} \al@gnfalse}
\def\tag#1{\relax \global\advance\eqnumber by1
       \ifs@cd
           \ifs@c
               \xdef#1{{\noexpand\rm(\the\secnumber .\the\eqnumber)}}
           \else
               \xdef#1{{\noexpand\rm(\char\the\appnumber .\the\eqnumber)}}
           \fi
       \else
           \xdef#1{{\noexpand\rm(\the\eqnumber)}}
       \fi
       \ifal@gn
           & #1
       \else
           \eqno #1
       \fi}
\def\eqn{\tag\?} 
\def\equ#1{#1} 
%
%
\newif\iffrontpage \frontpagefalse
\newif\ifletterstyle
\newif\ifmemostyle
\newif\ifpoemstyle
\newif\ifspanish \spanishfalse
\newtoks\memofootline
\newtoks\memoheadline
\newtoks\paperfootline
\newtoks\letterfootline
\newtoks\paperheadline
\newtoks\letterheadline
\newtoks\date
\newtoks\fecha
\newtoks\d@te
\headline={\ifmemostyle\the\memoheadline%
           \else \ifletterstyle\the\letterheadline%
           \else \ifpoemstyle\hfil%
           \else \the\paperheadline\fi\fi\fi}
\paperheadline={\hfil}
\letterheadline={\ifnum\pageno=1 \hfil
  \else\ifspanish\rm p\'agina \ \folio\hfil\the\fecha
  \else\rm page \ \folio\hfil\the\date\fi\fi}
\memoheadline={\ss Page \ \folio\hfil\the\date}
\footline={\ifmemostyle\the\memofootline\else\ifpoemstyle\hfil\else%
           \ifletterstyle\the\letterfootline\else\the\paperfootline\fi\fi\fi}
\letterfootline={\hfil}
\memofootline={\hfil}
\paperfootline={\iffrontpage\hfil\else \hss\iftwelv@\twelverm\else\tenrm\fi
-- \folio\ --\hss \fi }
\def\monthname{\relax\ifcase\month 0/\or January\or February\or
   March\or April\or May\or June\or July\or August\or September\or
   October\or November\or December\else\number\month/\fi}
\def\today{\monthname\ \number\day, \number\year}
\date={\today}
\def\nombremes{\relax\ifcase\month 0/\or Enero\or Febrero\or
   Marzo\or Abril\or Mayo\or Junio\or Julio\or Agosto\or Septiembre\or
   Octubre\or Noviembre\or Diciembre\else\number\month/\fi}
\def\hoy{\number\day\ de\ \nombremes, \number\year}
\fecha={\hoy}
\def\itpletterhead{
       \def\k{\kern -3pt}
       \font\sblogo=cmbx12 scaled \magstep5
       \setbox1=\hbox{\sblogo S\k t\k o\k n\k y\k B\k r\k o\k o\k k\k}
       \setbox2=\hbox{\vbox{
                      \hbox{\tenrm Jos\'e M. Figueroa-O'Farrill}
                      \hbox{\tenrm Institute for Theoretical Physics}
                      \hbox{\tenrm State University of New York at Stony Brook}
                      \hbox{\tenrm Stony Brook, NY\ \ 11794--3840}
                      \hbox{\tenrm telephone: (516) 632-7965}}}
\hbox to \hsize{\vbox to 1.2in{\vfill\box1\vfill}\hfill
                \vbox to 1.2in{\box2\vfill}}}
\hyphenation{U-ni-ver-si-teit u-ni-ver-si-teit The-o-re-tische the-o-re-tische
Fy-si-ca fy-si-ca}
\def\kulletterhead{
       \setbox1=\hbox{\vbox{
                      \hbox{\tenrm Jos\'e M. Figueroa-O'Farrill}
                      \hbox{\tenrm Instituut voor Theoretische Fysica}
                      \hbox{\tenrm Katholieke Universiteit Leuven}
                      \hbox{\tenrm Celestijnenlaan 200 D}
                      \hbox{\tenrm B-3001 Heverlee, BELGIUM}
                      \hbox{\tenrm e-mail: fgbda11@blekul11.bitnet}
                      \hbox{\tenrm telephone: (016) 201015 x3219}}}
\hbox to \hsize{\hfill\vbox to 1.5truein{\box1\vfill}}}
\def\bonnletterhead{
       \setbox1=\hbox{\vbox{
                      \hbox{\tenrm Jos\'e M. Figueroa-O'Farrill}
                      \hbox{\tenrm Physikalisches Institut der}
                      \hbox{\tenrm Universit\"at Bonn}
                      \hbox{\tenrm Nussallee 12}
                      \hbox{\tenrm W-5300 Bonn 1, GERMANY}
                      \hbox{\tenrm e-mail: ?}
                      \hbox{\tenrm telephone: (0228)?}}}
\hbox to \hsize{\hfill\vbox to 1.5truein{\box1\vfill}}}
\def\jsletterhead{
       \setbox1=\hbox{\vbox{
                      \hbox{\tenrm Jos\'e M. Figueroa-O'Farrill}
                      \hbox{\tenrm Stany Schrans}
                      \hbox{\tenrm Instituut voor Theoretische Fysica}
                      \hbox{\tenrm Katholieke Universiteit Leuven}
                      \hbox{\tenrm Celestijnenlaan 200 D}
                      \hbox{\tenrm B-3001 Heverlee, BELGIUM}
                      \hbox{\tenrm e-mail: fgbda31@blekul11.bitnet}}}
\hbox to \hsize{\hfill\vbox to 1.5truein{\box1\vfill}}}
\def\jeletterhead{
       \setbox1=\hbox{\vbox{
                      \hbox{\tenrm Jos\'e M. Figueroa-O'Farrill}
                      \hbox{\tenrm Eduardo Ramos}
                      \hbox{\tenrm Instituut voor Theoretische Fysica}
                      \hbox{\tenrm Katholieke Universiteit Leuven}
                      \hbox{\tenrm Celestijnenlaan 200 D}
                      \hbox{\tenrm B-3001 Heverlee, BELGIUM}
                      \hbox{\tenrm e-mail: fgbda11@blekul11.bitnet}}}
\hbox to \hsize{\hfill\vbox to 1.5truein{\box1\vfill}}}
\def\Date{\ifspanish\d@te=\fecha\else\d@te=\date\fi
\line{\hfill\rm\the\d@te}\bigskip}

%

%
\def\paperstyle{\letterstylefalse\normalspace\papersize}
\def\letterstyle{\letterstyletrue\singlespace\lettersize\parindent=0pt
                 \advance\parskip by 2\parskip}
\def\storystyle{\letterstyletrue\singlespace\lettersize
                 \advance\parskip by 1.5\parskip}

\def\papersize{\hsize=14 truecm\vsize=22 truecm\hoffset=6.5truemm
               \skip\footins=\bigskipamount}
\def\lettersize{\hsize=14truecm\vsize=22truecm\hoffset=6.5truemm
   \skip\footins=\smallskipamount \multiply\skip\footins by 3 }
\paperstyle   
%
%
%
\newskip\frontpageskip
\newif\ifp@bblock \p@bblocktrue
\newif\ifm@nth \m@nthtrue
\newtoks\pubnum
\newtoks\pubtype
\newtoks\m@nthn@me
\newcount\Ye@r
\advance\Ye@r by \year
\advance\Ye@r by -1900
\def\Year#1{\Ye@r=#1}
\def\Month#1{\m@nthfalse \m@nthn@me={#1}}
\def\m@nthname{\ifm@nth\monthname\else\the\m@nthn@me\fi}
\def\titlepage{\global\frontpagetrue\paperstyle\hrule height\z@ \relax
               \ifp@bblock\pubblock\fi\relax }
\def\endtitlepage{\vfil\break
                  \frontpagefalse} 
\frontpageskip=12pt plus .5fil minus 2pt
\pubtype={\iftwelv@\twelvesl\else\tensl\fi\ (Preliminary Version)}
\pubnum={?}
\def\nopubblock{\p@bblockfalse}
\def\pubblock{\line{\hfil\iftwelv@\twelverm\else\tenrm\fi%
Preprint--KUL--TF--\number\Ye@r/\the\pubnum\the\pubtype}
              \line{\hfil\iftwelv@\twelverm\else\tenrm\fi%
\m@nthname\ \number\year}}
\def\title#1{\vskip\frontpageskip\Titlestyle{\caps #1}\vskip3\headskip}
\def\author#1{\vskip.5\frontpageskip\titlestyle{\caps #1}\nobreak}

\def\address#1{\par\kern 5pt\titlestyle{
\it #1}}
\def\andaddress{\par\kern 5pt \centerline{\sl and} \address}

\def\UT{\address{Department of Mathematics, 61200\break
                 University of Texas\break
                 Austin, Texas~78712, U.~S.~A.}}
\def\abstract#1{\par\dimen@=\prevdepth \hrule height\z@ \prevdepth=\dimen@
   \vskip\frontpageskip\spacecheck\sectionminspace
   \centerline{\iftwelv@\fourteencp\else\twelvecp\fi ABSTRACT}\vskip\headskip
   {\noindent #1}}
%

%
%
%
\def\leaderfill{\leaders\hbox to 1em{\hss.\hss}\hfill}
\def\boxit#1{\vcenter{\hrule\hbox{\vrule\kern8pt
      \vbox{\kern8pt#1\kern8pt}\kern8pt\vrule}\hrule}}

%
%
%
\def\ref#1{{\bf [#1]}}
\def\ie{{\it i.e.\/}}
\def\eg{{\it e.g.\/}}
\def\nl{\hfil\break}
\def\half{{1\over 2}}
%
%
%
%
%
\newif\ifm@thstyle \m@thstylefalse
\def\mathstyle{\m@thstyletrue}
\def\proclaim#1#2\par{\smallbreak\begingroup
\advance\baselineskip by -0.25\baselineskip%
\advance\belowdisplayskip by -0.35\belowdisplayskip%
\advance\abovedisplayskip by -0.35\abovedisplayskip%
    \noindent{\bf#1.\enspace}{#2}\par\endgroup%
\smallbreak}
\def\m@kem@th#1#2#3{%
   \ifm@thstyle \global\advance\eqnumber by1
     \ifs@cd
         \ifs@c
           \xdef #1{{\noexpand #2\ \the\secnumber .\the\eqnumber}}
         \else
           \xdef #1{{\noexpand #2\ \char\the\appnumber .\the\eqnumber}}
         \fi
     \else
         \xdef #1{{\noexpand #2\ \the\eqnumber}}
     \fi
     \proclaim{#1}{#3}
   \else
     \proclaim{#2}{#3}
   \fi}
%
\def\Thmtag#1#2{\m@kem@th #1{Theorem}{\sl#2}}
\def\Proptag#1#2{\m@kem@th #1{Proposition}{\sl#2}}
\def\Deftag#1#2{\m@kem@th #1{Definition}{\rm#2}}
\def\Lemtag#1#2{\m@kem@th #1{Lemma}{\sl#2}}
\def\Cortag#1#2{\m@kem@th #1{Corollary}{\sl#2}}
\def\Conjtag#1#2{\m@kem@th #1{Conjecture}{\sl#2}}
\def\Rmktag#1#2{\m@kem@th #1{Remark}{\rm#2}}
\def\Exmtag#1#2{\m@kem@th #1{Example}{\rm#2}}
\def\Qrytag#1#2{\m@kem@th #1{Query}{\it#2}}
\def\Thm#1{\m@kem@th\?{Theorem}{\sl#1}}
\def\Prop#1{\m@kem@th\?{Proposition}{\sl#1}}
\def\Def#1{\m@kem@th\?{Definition}{\rm#1}}
\def\Lem#1{\m@kem@th\?{Lemma}{\sl#1}}
\def\Cor#1{\m@kem@th\?{Corollary}{\sl#1}}
\def\Conj#1{\m@kem@th\?{Conjecture}{\sl#1}}
\def\Rmk#1{\m@kem@th\?{Remark}{\rm#1}}
\def\Exm#1{\m@kem@th\?{Example}{\rm#1}}
\def\Qry#1{\m@kem@th\?{Query}{\it#1}}
\let\Pf=\Proof

%
%
\def\qed{\vrule width 0.7em height 0.6em depth 0.2em}
\def\implies{\Rightarrow}
\def\lapprox{\hbox{\lower3pt\hbox{$\buildrel<\over\sim$}}}
\def\gapprox{\hbox{\lower3pt\hbox{$\buildrel<\over\sim$}}}
\def\quotient#1#2{#1/\lower0pt\hbox{${#2}$}}
%
%
\def\into{\hookrightarrow}
\def\to{\rightarrow}
\def\tto{\longrightarrow}
\def\isomap{\buildrel \cong \over \tto}
%
\def\mapright#1{\smash{
    \mathop{\tto}\limits^{#1}}}

%
%
%
\def\reals{{\bf R}} 
\def\integ{{\bf Z}} 
%
%
\def\underrightarrow#1{\vtop{\ialign{##\crcr
      $\hfil\displaystyle{#1}\hfil$\crcr
      \noalign{\kern-\p@\nointerlineskip}
      \rightarrowfill\crcr}}} 
\def\underleftarrow#1{\vtop{\ialign{##\crcr
      $\hfil\displaystyle{#1}\hfil$\crcr
      \noalign{\kern-\p@\nointerlineskip}
      \leftarrowfill\crcr}}}  

%
%
\def\comm#1#2{\bigl[#1\, ,\,#2\bigr]}
\def\stc#1#2#3{{f_{#2#3}}^{#1}}
%
%
\def\lied#1#2{{\cal L}_{#1}{#2}}
%
%
%
%
%
%
\newdimen\unit
\newdimen\redunit
%
%
\def\p@int#1:#2 #3 {\rlap{\kern#2\unit
     \raise#3\unit\hbox{#1}}}
%
%
\def\th@r{\vrule height0\unit depth.1\unit width1\unit}
\def\bh@r{\vrule height.1\unit depth0\unit width1\unit}
\def\lv@r{\vrule height1\unit depth0\unit width.1\unit}
\def\rv@r{\vrule height1\unit depth0\unit width.1\unit}
%
%
\def\t@ble@u{\hbox{\p@int\bh@r:0 0
                   \p@int\lv@r:0 0
                   \p@int\rv@r:.9 0
                   \p@int\th@r:0 1
                   }
             }
%
%
\def\t@bleau#1#2{\rlap{\kern#1\redunit
     \raise#2\redunit\t@ble@u}}
%
%
\newcount\n
\newcount\m
\def\makecol#1#2#3{\n=0 \m=#3
  \loop\ifnum\n<#1{}\advance\m by -1 \t@bleau{#2}{\number\m}\advance\n by 1
\repeat}
%
%
\def\makerow#1#2#3{\n=0 \m=#3
 \loop\ifnum\n<#1{}\advance\m by 1 \t@bleau{\number\m}{#2}\advance\n by 1
\repeat}
%
%
\def\checkunits{\ifinner \unit=6pt \else \unit=8pt \fi
                \redunit=0.9\unit } 
\def\ytsym#1{\checkunits\kern-.5\unit
  \vcenter{\hbox{\makerow{#1}{0}{0}\kern#1\unit}}\kern.5em} 
\def\ytant#1{\checkunits\kern.5em
  \vcenter{\hbox{\makecol{#1}{0}{0}\kern1\unit}}\kern.5em} 
\def\ytwo#1#2{\checkunits
  \vcenter{\hbox{\makecol{#1}{0}{0}\makecol{#2}{1}{0}\kern2\unit}}
                  \ } 
\def\ythree#1#2#3{\checkunits
  \vcenter{\hbox{\makecol{#1}{0}{0}\makecol{#2}{1}{0}\makecol{#3}{2}{0}%
\kern3\unit}}
                  \ } 
%
%
%

%
%
\catcode`\@=12 
\def\papersize{\hsize=17 truecm\vsize=22 truecm\hoffset=0truemm
               \skip\footins=\bigskipamount}

\def\utletterhead{
       \setbox1=\hbox{\vbox{
                      \hbox{\tenrm Takashi Kimura}
                      \hbox{\tenrm Department of Mathematics, 61200}
                      \hbox{\tenrm University of Texas}
                      \hbox{\tenrm Austin, Texas 78712, U.~S.~A.}
                      \hbox{\tenrm e-mail: kimura@math.utexas.edu}
                      \hbox{\tenrm telephone: (512)471-0122}}}
\hbox to \hsize{\hfill\vbox to 1.5truein{\box1\vfill}}}
\def\UTletterhead{
       \setbox1=\hbox{\vbox{
                      \hbox{Takashi Kimura}
                      \hbox{Department of Mathematics, 61200}
                      \hbox{University of Texas}
                      \hbox{Austin, Texas 78712, U.~S.~A.}
                      \hbox{e-mail: kimura@math.utexas.edu}
                      \hbox{telephone: (512)471-0122}}}
\hbox to \hsize{\hfill\vbox to 1truein{\box1\vfill}}}
\def\abs#1{{|{#1}|}}				
\def\Cinf{C^\infty}				
\def\cl#1{\lbrack\!\lbrack\,{#1}\,\rbrack\!\rbrack} 
\def\comm#1#2{[{#1}\,,\,{#2}]}			
\def\Dsum{\bigoplus}    	                
\def\dsum{\oplus}				
\def\exterior{\Lambda}				
\def\interior#1{\iota({#1})}			
\def\intersect{\,\cap\,}			
\def\mod{\ {\rm mod}\ }				
\def\sec#1{\Gamma({#1})}			
\def\smooth#1{C^\infty({#1})}			
\def\symmetric{{\cal S}}			

%
%
\def\Ann{{\rm Ann}}				
\def\b#1#2{b^{({#1})}_{#2}}			
\def\C{{\cal C}}				
\def\c#1#2{c^{#2}_{({#1})}}			
\def\cltwo#1{\langle\,{#1}\,\rangle}		
\def\clthree#1{\cl{#1}}				
\def\D#1{{\,D^{({#1})}\,}}			
\def\Dist{V}					
\def\d#1#2{{\delta^{(#1)}_{#2}}}		
\def\dd{\delta}					
\def\dk#1{\,{\delta^{(#1)}}\,}			
\def\db{{\overline{\delta}}}			

\def\Evertical#1{\exterior_{\smooth{M}/{#1}}({#1}/{#1}^2)}
\def\Everticalp#1#2{\exterior_{\smooth{M}/{#2}}^{#1}({#2}/{#2}^2)}
\def\Evert#1{\exterior_{\smooth{M}/{#1}}({#1}'/{#1}^2)}
\def\Evertp#1#2{\exterior_{\smooth{M}/{#2}}^{#1}({#2}'/{#2}^2)}
\def\F{{\cal F}}				
\def\gammabar{\overline{\gamma}}		
\def\K#1{{K_{(#1)}}}				
\def\KT#1{{\cal K}^{(#1)}}	 		
\def\M#1#2{M_{{(#1)}{#2}}}			
\def\Mp#1#2{M^\prime_{{(#1)}{#2}}}		
\def\Mtilde{\widetilde{M}} 			
\def\N{{\cal N}}				
\def\Ntilde{\widetilde{N}}			
\def\Omegatilde{\widetilde{\Omega}} 		
\def\P{{\cal P}}				
\def\Psh{P_\#}					
\def\Ptilde{\widetilde{P}}			
\def\r{\mu}					
\def\SS#1{\Symmetric_{\smooth{M}}({#1})}	
\def\Symmetric{\symmetric}			
\def\U{{\cal U}}				
\def\V#1{{{\bf V}^{(#1)}}}			
\def\Z#1#2#3{\,{Z^{(#1)}\,}_{#2}^{#3}\,}	

\mathstyle
\def\pubblock{ \line{\hfil\twelverm July 1992}}
\titlepage
\title{GENERALIZED CLASSICAL BRST COHOMOLOGY AND REDUCTION OF POISSON
MANIFOLDS}
\author{Takashi Kimura\footnote{${}^\clubsuit$}{\twelverm Internet:
kimura@math.utexas.edu.  Address after September 1992: Department of\nl
\message{Unnatural page break} Mathematics, University of Pennsylvania}}
\UT
\abstract{In this paper, we formulate a generalization of the classical BRST
construction which applies to the case of the reduction of a poisson manifold
by a submanifold.  In the case of symplectic reduction, our procedure
generalizes the usual classical BRST construction which only applies
to symplectic reduction of a symplectic manifold by a coisotropic
submanifold, \ie\ the case of reducible ``first class'' constraints.  In
particular, our procedure yields a method to deal with ``second-class''
constraints.  We construct the BRST complex and compute its cohomology.  BRST
cohomology vanishes for negative dimension and is isomorphic as a poisson
algebra to the algebra of smooth functions on the reduced poisson manifold
in zero dimension.  We then show that in the general case of reduction of
poisson manifolds, BRST cohomology cannot be identified with the cohomology
of vertical differential forms.}
\par\nobreak\vfil\nobreak\medskip\centerline{To Appear {\sl Communications
in Mathematical Physics}}
\endtitlepage

\section{Introduction}

Classical BRST cohomology has a long history in the physics
literature, \eg\ ${\Ref\Physics}$. Although its origins are in the
context of quantum field theory,  it is now known that classical BRST
cohomology is a cohomology theory that contains all of the information
of the symplectic reduction of a symplectic manifold by a closed and
embedded coisotropic submanifold.$^{{\Ref\FHST}, {\Ref\Stasheff}}$ In
the language of Dirac $^{\Ref\Dirac}$, this corresponds to symplectic
reduction arising from (possibly reducible) ``first class constraints.''
The classical BRST complex is constructed using only purely algebraic
properties of the poisson algebra of smooth functions on the original
(unreduced) symplectic manifold and some of its ideals.  Furthermore,
since the classical BRST complex is a poisson superalgebra and the
differential a poisson derivation, classical BRST cohomology inherits
the structure of a poisson superalgebra.  Classical BRST cohomology
is isomorphic (as poisson algebras) in zero dimension to the algebra
of smooth functions on the symplectic reduction.$^{\Ref\FHST}$  When
the (symplectic) normal bundle of the coisotropic submanifold is a
trivial bundle, classical BRST cohomology in nonnegative dimension is
isomorphic to the cohomology of vertical differential forms with
respect to the null foliation.$^{\Ref\DuVi}$ The results in
${\Ref\FHST}$ and ${\Ref\Stasheff}$ suggest that this is the case
even if the normal bundle is not a trivial bundle.

In this paper, we transcribe the procedure of the reduction of a poisson
manifold $(M, P)$ by a closed and embedded submanifold into the language of
poisson algebras.   Inspired by this example, we give an algebraic definition
of the reduction of a poisson algebra by an ideal.  In the case of a
poisson manifold,  our algebraic definition gives rise to a notion of the
reduction of $\smooth{M}$ by any ideal, whether this ideal arises as the
ideal of functions which vanish on a closed and embedded submanifold or not.
In particular, we show that for certain special ideals (``coisotropic
ideals''), we can generalize classical BRST cohomology. Our construction
has some important applications.  Consider the case of the reduction of a
symplectic manifold by an arbitrary closed and embedded submanifold.
In the language of Dirac, the ideal of functions which vanish on this
submanifold $I$ is generated by a collection of  first class
constraints and second class constraints. Our method tells us how to
construct the classical BRST complex in this general case.  The idea
is to replace the original collection of constraints by a new set of
``first class constraints.'' Although this new set of constraints is
guaranteed to exist, our method does not explicitly construct them, in
general. However, in the case where the submanifold of the symplectic
manifold is itself a symplectic manifold (\ie\ $I$ is generated by
only second class constraints), there is a method to explicitly
construct the new set of constraints which would, in applications to
field theory, preserve any Lorentz covariance type properties of the
original collection of constraints.  An interesting consequence of our
generalization is that classical BRST cohomology is not generally
isomorphic to  the cohomology of vertical differential forms although
it is isomorphic as poisson algebras to the algebra of smooth
functions on the reduced poisson manifold in zero dimension.

This paper is organized as follows.  In section $2$, we transcribe the
procedure of reduction of a poisson manifold by a submanifold into purely
poisson algebraic terms.  In section $3$, we review the Koszul-Tate
resolution. In section $4$, we show that in the special case of symplectic
reduction by a symplectic submanifold, our procedure results in infinitely
reducible constraints. In section $5$, we construct the space of BRST
cochains and show that it forms a poisson superalgebra.  In section $6$,
we construct the BRST charge inductively.  In section $7$ we compute the
cohomology explicitly.  In section $8$, we explain why BRST cohomology
is not vertical cohomology.  Finally, section $9$ contains some
concluding remarks as well as some possible avenues for future research.

\ack

It is a pleasure to thank Don Wilbour for discussions, Jim Stasheff for his
careful reading the original manuscript, and Jos\`e Figueroa-O'Farrill for
the encouragement and for bending his ear in my direction.

\section {Reduction of Poisson Manifolds}

In this section, we review the reduction of a poisson manifold by a
submanifold. It is a procedure which becomes, in the case where the poisson
manifold is a symplectic manifold, symplectic reduction by a submanifold.
This reduction is done by transcription of this geometric procedure into the
language of poisson algebras (see ${\Ref\DuVi}$ and ${\Ref\WA}$).  The
algebraic formulation generalizes the geometric one since it can be applied
to cases where the reduced poisson manifold is not smooth.

Let $(M,P)$ be a $m$-dimensional poisson manifold, \ie\  $M$ is a smooth
$m$-dimensional manifold and $P$ is a bivector in $\exterior^2(TM)$ such that
the Schouten bracket of $P$ with itself vanishes.  Given a poisson manifold,
the {\sl poisson bracket} of two smooth functions on $M$, $f$ and
$g$, is given by
$$\comm{f}{g} = P(df,dg).\ \tag\BRACKET$$
Since $P$ is in $\exterior^2(TM)$, the poisson bracket is antisymmetric.
Furthermore, since the exterior derivative acts like a derivation on
$\smooth{M}$, we have
$$\comm{f}{gh} = \comm{f}{g}h+g\comm{f}{h}\ \tag\DERIVATION$$
for all $f,g,h$ in $\smooth{M}$.  Finally, the fact that the Schouten
bracket of $P$ with itself vanishes is equivalent to the Jacobi identity, \ie
$$\comm{f}{\comm{g}{h}} = \comm{\comm{f}{g}}{h} + \comm{g}{\comm{f}{h}}\
\tag\JACOBI$$
for all $f,g,h$ in $\smooth{M}$.  In other words, $\smooth{M}$ forms a
{\sl poisson algebra}, \ie\  $\smooth{M}$ is an associative and
commutative algebra with unit with respect to pointwise multiplication,
a Lie algebra with respect to the poisson bracket, and the two operations
intertwine via equation $\DERIVATION$. The poisson algebra $\smooth{M}$
completely characterizes the poisson manifold $(M,P)$.  Furthermore, all
poisson structures on $\smooth{M}$ arise from endowing $M$ with a suitable
poisson structure.

Given a poisson manifold $(M,P)$ , there is a map $\Psh:T^*M\,\to\,TM$ given
by
$$\Psh\alpha = \interior{\alpha}P \eqn$$
for all $\alpha$ in $T_m^*M$ and points $m$ in $M$ where $\interior{\alpha}$
is the interior product.  $\Psh$ allows us to define the {\sl hamiltonian
vector field associated to a function $f$ in $\smooth{M}$} by $X_f =
\Psh(df)$. In  terms of poisson algebras, this definition is equivalent to
$X_f = \comm{f}{\cdot}$ since equation $\DERIVATION$ insures that
$\comm{f}{\cdot}$ is a derivation with respect to pointwise multiplication in
$\smooth{M}$ and, hence, a vector field on $M$.  Furthermore, equation
$\JACOBI$ implies that
$$X_{\comm{f}{g}} = \comm{X_f}{X_g}\ \eqn$$
for all $f,g$ in $\smooth{M}$ where the bracket on the right hand side is the
Lie bracket.  In other words, the map $f\,\mapsto\,X_f$ is a Lie algebra
homomorphism from $\smooth{M}\,\to\,\sec{TM}$.

Consider the closed and embedded submanifold $i:M_o\,\into\,(M,P)$ which has
codimension $k$. The submanifold $M_o$ is completely characterized by its
associated algebra of smooth functions $\smooth{M_o}$. Let us denote the
ideal of functions in $\smooth{M}$  which vanish on $M_o$ by $I$.  Since
$M_o$ is a closed and embedded submanifold of $M$, we have the isomorphism of
(associative) algebras
$${{\smooth{M}}\over I}\isomap\smooth{M_o}\ \tag\FUNCTIONS$$
given by $\cl{f}\,\mapsto\,f|_{M_o}$.  This map is certainly well-defined.
It is injective since the elements in $\smooth{M}/I$ which give rise to the
zero map on $M_o$ are those which vanish on $M_o$ and it is surjective since
any smooth function on  $M_o$ arises from the restriction of some smooth
function on $M$.  We can do a similar construction with vector fields. Any
vector field on $M_o$ arises as the restriction of some vector field on $M$.
However, the restriction of a vector field on $M$ to $M_o$ is not, in
general, a vector field on $M_o$ since it need not be tangent to $M_o$.  The
space of all vector fields on $M$ which restricts to a vector field on $M_o$
is given by
$$ \N(I) = \{X\in\sec{TM}\,|\,X(i)\in I\ \forall\,i\in I\}.\ \eqn$$
Of course, two vector fields in $\N(I)$ may restrict to the same vector field
on $M_o$.  This happens only if their difference vanishes on $M_o$.  Since
vector fields in $\N(I)$ which vanish on $M_o$ are precisely the elements in
$I\N(I)$, we have the isomorphism of Lie algebras and of
$\smooth{M}/I\,\cong\, \smooth{M_o}$-modules
\message{lie alg. iso and assoc iso -> Rinehart modules?}
$${{\N(I)}\over {I\N(I)}} \isomap \sec{TM_o}\ \tag\VECTORFIELDS$$
via the map $\cl{X}\,\mapsto\, X|_{M_o}$.  Equation $\VECTORFIELDS$ has an
algebraic interpretation, as well.  Vector fields on $M$ and $M_o$ can be
identified with the derivations on $\smooth{M}$ and $\smooth{M_o}$,
respectively. The isomorphism $\smooth{M_o}\isomap\smooth{M}/I$ tells us that
derivations on $\smooth{M_o}$ should be induced from derivations on
$\smooth{M}$. The Lie subalgebra $\N(I)$ consists of precisely those
derivations in $\smooth{M}$ which respect the ideal $I$ and, hence, induce
derivations on $\smooth{M}/I$. The ideal $I\N(I)$ consists of those
derivations in $\N(I)$ which induce the zero map on $\smooth{M}/I$.

The hamiltonian vector fields on $M$ are just the inner derivations of
$\smooth{M}$.  It is natural to ask when a hamiltonian vector field restricts
to a vector field on $M_o$.  Suppose that $X_f$ restricts to a vector field
on $M_o$ then $X_f$ must belong to $\N(I)$.  This means that
$X_f(i)=\comm{f}{i}$ must belong to $I$ for all $i$ in $I$.  In other words,
functions whose hamiltonian vector fields when restricted to $M_o$ are vector
fields on $M_o$ are those functions in the {\sl normalizer of} $I$ denoted by
$$N(I) = \{f\in\smooth{M}\,|\,\comm{f}{i}\in I\ \forall\,i\in I\}.\ \eqn$$
Notice that $N(I)$ forms a poisson subalgebra of $\smooth{M}$.  It will play
an important role in what follows.

Under certain conditions, $M_o$ has an associated involutive distribution
such that the space of leaves of its associated foliation inherits the
structure of a poisson manifold.  Let us describe this situation in more
detail.

Denote the pullback of $TM$ and $T^*M$  to $M_o$ via the inclusion map
by $i^{-1}TM$ and $i^{-1}T^*M$, respectively.  The poisson structure $P$ on
$M$ can be pulled back via the inclusion map to an element in
$\Lambda^2(i^{-1}T^*M)$ which we shall also denote by $P$ to avoid notational
clutter.  It allows us to define a rank $k$ subbundle of $i^{-1}T^*M$ whose
fibers consist of $1$-forms which  vanish when evaluated on vectors tangent
to $M_o$ called the {\sl annihilator bundle} (or the {\sl conormal bundle})
of $M_o$.  It is denoted by $\Ann(TM_o)\,\to\,M_o$ and its fibers are given
by
$$\Ann_m(TM_o) = \{\alpha\in i^{-1}T_m^*M\,|\,\alpha(v) = 0\ \forall\,v\in
T_mM_o\}\ \eqn$$
for all points $m$ in $M_o$.  At every point $m$ in $M_o$, let us define
$$ T_mM_o^\perp = \Psh\Ann_m(TM_o) =
\{\Psh\alpha\,|\,\alpha\in\Ann_m(TM_o)\}.\ \eqn$$
Let us assume that $TM_o^\perp$ has constant rank so that $TM_o^\perp$ forms
a subbundle of $i^{-1}TM$. The {\sl null distribution} of $M_o$ is
given by $m\,\mapsto\,\Dist_m$ for all $m$ in $M_o$ where
$$ \Dist_m = T_mM_o\intersect T_mM_o^\perp.\eqn$$
Let us assume that $\Dist$ has constant rank so that $\Dist$ forms a
subbundle of $TM_o$.  We will show that the null distribution is an
involutive distribution over $M_o$.

Let us begin by recalling several facts.  First of all, given any $i$
in $I$, $di|_{M_o}$ belongs to $\sec{\Ann(TM_o)}$ since for all
$\cl{v}\in\N(I)/I\N(I)$, $di(v) = v(i)$ which belongs to $I$ by the
definition of $\N(I)$ and, therefore, vanishes when restricted to $M_o$.
Furthermore, the exterior derivative of an element in $I^2$ always vanishes
when restricted to $M_o$. Therefore, we have the well-defined map
$$I/I^2 \,\isomap\, \sec{\Ann(TM_o)}\tag\ANNISO$$
given by $\cl{i}\,\mapsto\, di|_{M_o}$.  This map is readily seen to be an
isomorphism by showing the isomorphism locally and then globalizing it by
using partitions of unity.   Thus, all sections of $TM_o^\perp$ are
restrictions to  $M_o$ of hamiltonian vector fields of elements in $I$.
Since the sections of the null distribution are the sections of
$TM_o\intersect TM_o^\perp$ and $N(I)$ consists of all functions on $M$ whose
hamiltonian vector fields when restricted to $M_o$ are vector fields on
$M_o$, all sections of $TM_o\intersect TM_o^\perp$ are restrictions to $M_o$
of hamiltonian vector fields of elements in
$$I' = N(I)\intersect I.\ \eqn$$
Notice that $I'$ is naturally a poisson subalgebra of $N(I)$ and, therefore,
$$\comm{X_{i_1}}{X_{i_2}} = X_{\comm{i_1}{i_2}}\ \eqn$$
for all $i_1,i_2$ in $I'$.  This proves that $\Dist$ is an involutive
distribution on $M_o$.

By the Frobenius theorem, associated to the involutive distribution $\Dist$
there exists a foliation of $M_o$ by maximal connected submanifolds (called
{\sl leaves}) such that the tangent space to each leaf is the restriction of
$\Dist$.  Let us denote the space of leaves of the foliation by $\Mtilde$.
Let us assume, furthermore, that  conditions are such that the projection map
$\pi:M_o\,\to\,\Mtilde$ which takes each point on $M_o$ and projects it into
the leaf containing it is a smooth map. \message{Does pi:M_o -> \Mtilde
smooth ==> Bundle?} In this case, we will show that $\Mtilde$ has an induced
poisson structure $\Ptilde$.  We will construct $\Ptilde$ by inducing a
poisson algebra structure on $\smooth{\Mtilde}$ from the poisson algebra
structure on $\smooth{M}$.

The functions in $\smooth{M}$ which induce smooth functions on $\Mtilde$ are
those which, when restricted to $M_o$ are constant along each leaf of the
foliation.  Since each leaf is connected and has tangent vector fields which
are restrictions of hamiltonian vector fields of elements in $I'$, a function
$f$ in $\smooth{M}$ induces a function in $\smooth{\Mtilde}$ if and only if $
\lied{X_i}f|_{M_o} = 0$ for all $i$ in $I'$. Equivalently, $f$ induces a
smooth function on $\Mtilde$ if and only if  $f$ belongs to $N(I',I)$ where
$$N(I',I) = \{f\in\smooth{M}\,|\,\comm{f}{i'}\in I,\ \forall\,i'\in I'\}.\
\eqn$$
Furthermore, all functions in $\smooth{\Mtilde}$ are induced from functions
in $N(I',I)$ since any smooth function $\widetilde{f}$ on $\Mtilde$,
can be extended to a smooth function $\pi^*\widetilde{f}$ on $M_o$ which
projects to it.  Since two functions in $N(I',I)$ restrict to the same
function on $M_o$  and, therefore, induce the same function on $\Mtilde$
if and only if they differ by an element of $I$, we have the isomorphism
of associative algebras
$$ \smooth{\Mtilde}\isomap {{N(I',I)}\over I}.\ \tag\ALMOSTREDUCTION$$
This is not obviously an isomorphism of poisson algebras since $N(I',I)/I$
does not appear to naturally inherit the structure of a poisson algebra from
$\smooth{M}$ as it stands.  In order to obtain a poisson algebra for the
right hand side of this equation, we need to delve deeper into the algebraic
structure of $N(I',I)$.

Suppose that $f$ is an element of $N(I',I)$, then for all $i'$ in $I'$,
$df(X_{i'}) = \comm{i'}{f}$ belongs to $I$ and, therefore, vanishes when
restricted to $M_o$.  In other words, we have
$$ df|_{M_o} \in \sec{\Ann(TM_o\intersect TM_o^\perp)}\ \eqn$$
since all sections of the null distribution are given by restrictions to
$M_o$ of hamiltonian vector fields of elements in $I'$. However, it is a
general fact from linear algebra that if $V$ is a (finite dimensional) vector
space and $W_1$ and $W_2$ are subspaces then
$$ \Ann(W_1\intersect W_2) = \Ann(W_1)+\Ann(W_2).\ \eqn$$
We can see this as follows. Consider $\alpha$ in $\Ann(W_1)$ and $\beta$ in
$\Ann(W_2)$ then $\alpha+\beta$ certainly lies in $\Ann(W_1\intersect W_2)$
so that $\Ann(W_1)+\Ann(W_2)\subseteq\Ann(W_1\intersect W_2)$. The equality
follows from some linear algebra and by counting dimensions. Therefore, we
obtain the result
$$df|_{M_o} \in\sec{\Ann(TM_o)}+\sec{\Ann(TM_o^\perp)}.\ \eqn$$
Since $\sec{\Ann(TM_o)}$ consists of the restriction to $M_o$ of the exterior
derivative of elements in $I$, there exists an $i$ in $I$ such that
$$ (df-di)|_{M_o} \in\sec{\Ann(TM_o^\perp)}.\ \tag\LASTEQ$$

Let us use another fact.  Consider $\alpha$ in $\Ann_m(TM_o^\perp) =
\Ann_m(\Psh(\Ann(TM_o)))$ where  $m$ is a point in $M_o$ then
$\alpha(\Psh\beta) = 0$ for all $\beta$ in $\Ann_m(TM_o)$.  However, $0 =
\alpha(\Psh\beta) = P(\beta,\alpha) = -P(\alpha,\beta) = -\beta(\Psh \alpha)$
for all $\beta$ in $\Ann_m(TM_o)$ means that $\Psh\alpha$ must belong to
$T_mM_o$.  In other words, $\Psh\Ann_m(TM_o^\perp)\subseteq T_mM_o$ for all
$m$ in $M_o$.  Applying $\Psh$ to both sides of equation $\LASTEQ$ yields
$$ X_{(f-i)}|_{M_o}\in\sec{TM_o}.\ \eqn$$
This tells us that $f-i$ belongs to $N(I)$. In other words, $f$ belongs to
$N(I)+I$.  Combining this result with $\ALMOSTREDUCTION$, we get
$$\smooth{\Mtilde}\isomap {{N(I)+I}\over I}.\ \tag\PRELAST$$
However, the right hand side is still not obviously a poisson algebra.  This
can be remedied by using a basic fact from linear algebra.  Given two
subspaces $W_1$ and $W_2$ of a vector space $V$, there is the canonical
isomorphism
$${{W_1+W_2}\over{W_2}} \isomap {{W_1}\over{W_1\intersect W_2}}\ \eqn$$
defined by $\cl{w_1+w_2}\,\mapsto\,\cl{w_1}$ for all $w_1$ in $W_1$ and $w_2$
in $W_2$.  Using this fact on the right hand side of equation $\PRELAST$, we
obtain the isomorphism $\smooth{\Mtilde}\isomap N(I)/I'$.  The right hand
side is naturally a poisson algebra since $N(I)$ is a poisson subalgebra of
$\smooth{M}$ and $I'$ is a {\sl poisson ideal of $N(I)$,} \ie\ $I'$ is an
ideal with respect to both the lie bracket and multiplication operations in
$N(I)$.  We use this isomorphism to endow $\smooth{\Mtilde}$ with the
structure of a poisson algebra thereby completing the process of inducing a
poisson structure $\Ptilde$ on $\Mtilde$ from the poisson manifold $(M,P)$.
Therefore, this isomorphism induces a poisson structure $\Ptilde$ on
$\Mtilde$.  We have just shown the following result.

\Thmtag\FUNCTIONS{Let $i:M_o\,\into\,(M,P)$ be a closed and embedded
submanifold of a poisson manifold. Assume that $TM_o^\perp$ and the
null distribution have constant rank. Furthermore, let us assume that the
canonical projection map of a point in $M_o$ into the leaf that contains it,
$\pi:M_o \,\to\,\Mtilde$, is a smooth map.  In this case, we have the
isomorphism
$$\smooth{\Mtilde}\isomap{{N(I)}\over {I'}}\ \eqn$$
where $I$ is the ideal of functions vanishing on $M_o$, $N(I)$ is the
normalizer of $I$ in $\smooth{M}$, and $I' = N(I)\intersect I$.  Since the
right hand side is naturally a  poisson algebra, the isomorphism defines a
poisson structure $\Ptilde$ on $\Mtilde$. The poisson manifold
$(\Mtilde,\Ptilde)$ is said to be the {\rm reduction of the the poisson
manifold $(M,P)$ by $M_o$.}}

This theorem shows that the process of reduction of a poisson manifold by a
submanifold is essentially an algebraic one.  This leads to the following
purely algebraic definition.

\Def{Let $\P$ be a poisson algebra, $J$ be an ideal of $\P$, $N(J)$ be the
normalizer of $J$ in $\P$, and $J' = N(J)\intersect J$.  We say that
the poisson algebra $N(J)/J'$ is the {\sl reduction of the poisson algebra
$\P$ by the ideal $J$}.}

The reduction of the poisson algebra of functions on a poisson manifold by
the ideal of functions which vanish on a submanifold is well-defined even if
the projection map $\pi:M_o\,\to\,\Mtilde$ is not smooth. Therefore, this
algebraic definition of reduction generalizes the geometric one.  Also,
notice that $I'$ given in the above definition is generally nonzero since
$$I^2\subseteq I'.\ \tag\IPRIME$$

Consider the special case where the poisson manifold $(M,P)$ is, in fact, a
symplectic manifold.  This occurs when $P$ is a nondegenerate bivector. Its
inverse, $\omega$, is a $2$-form on $M$ which is closed because the Schouten
bracket of $P$ with itself vanishes thereby making  $(M,\omega)$ into a
symplectic manifold.  Let us call the poisson algebra $\smooth{M}$ a {\sl
symplectic algebra} if $(M,P)$ is a symplectic manifold.  There is, in fact,
an algebraic characterization of this fact, \ie\   $\smooth{M}$ is a
symplectic algebra if and only if the kernel of the map from
$\smooth{M}\,\to\,\sec{TM}$ given by $f\,\mapsto\,\comm{f}{\cdot}$ consists
of the locally constant functions.

The symplectic reduction of $(M,\omega)$ by the submanifold $M_o$
is precisely the procedure of the reduction of the poisson manifold $(M,P)$
by $M_o$.  Notice that here $TM_o^\perp$ is the usual symplectic normal
bundle to $TM_o$ and the null distribution on $M_o$ arises as the null
space of the pullback via the inclusion map of $\omega$ to $M_o$. There are
two extremes to symplectic reduction.  The first is when $M_o$ is a
coisotropic submanifold of $(M,\omega)$, \ie\   $T_mM_o^\perp\subseteq
T_mM_o$ for all $m$ in $M_o$.  In this case, the null distribution is just
$TM_o^\perp$. Let $I$ be the ideal of functions in $\smooth{M}$ which vanish
on $M_o$.  If $M_o$ is coisotropic, sections of $TM_o^\perp$ belong to the
space of sections of $TM_o$. However, all sections of $TM_o^\perp$ are
restrictions of hamiltonian vector fields of elements in $I$ to $M_o$.  Since
$N(I)$ consists of all functions on $M$ whose hamiltonian vector fields when
restricted to $M_o$ are sections of $TM_o$, we have $I\subseteq N(I)$. This
is equivalent to the statement that the ideal $I$ forms a poisson subalgebra
of $\smooth{M}$, \ie\
$$ \comm{I}{I}\subseteq I.\ \tag\COISOTROPIC$$
It is clear that $M_o$ is a coisotropic submanifold if and only if equation
$\COISOTROPIC$ holds. In this case, the symplectic reduction of $(M,P)$ by
$M_o$, $(\Mtilde,\Ptilde)$, is algebraically given by
$$ \smooth{\Mtilde} \isomap {{N(I)}\over I}.\ \eqn$$
This geometric procedure inspires the following algebraic definition.

\Def{Let $\P$ be a poisson algebra, $J$ an ideal, and $N(J)$ the normalizer
of $J$ in $\P$.  The ideal $J$ is said to be a {\sl coisotropic ideal of
$\P$} if and only if $J$ is a poisson subalgebra of $\P$.}

Clearly, if $J$ is an ideal of $\smooth{M}$ then $J'$ is a coisotropic ideal
of $\smooth{M}$.

The other extreme occurs when $M_o$ is a symplectic submanifold of
$(M,\omega)$. In this case, since $i^*\omega$ is already nondegenerate, the
null distribution vanishes, \ie\  $T_mM_o\intersect T_mM_o^\perp =
0$.  However, $\sec{TM_o\intersect TM_o^\perp}$ are hamiltonian vector fields
of elements in $I'$ restricted to $M_o$, therefore,  for all $i$ in $I'$ we
have $X_i|_{M_o} = 0$. However, $X_i = \Psh di$ and $\Psh$ is an isomorphism
since $(M,P)$ is a symplectic manifold.  Therefore, we have $di|_{M_o} = 0$
for all $i$ in $I'$.  However, the elements in $I$ which satisfy $di|_{M_o} =
0$ are just the elements in $I^2$, therefore, $I'\subseteq I^2$. Combining
this with equation $\IPRIME$, we conclude that $M_o$ is a symplectic
submanifold of the symplectic manifold $(M,P)$ if and only if
$$ I' = I^2.\ \eqn$$
Therefore, if $(\Mtilde, \Omegatilde)$ is the symplectic reduction of
$(M,\omega)$ by the symplectic submanifold $M_o$ then we conclude that
$$ \smooth{\Mtilde} \isomap {{N(I)}\over I^2}.\ \eqn$$

Classical BRST cohomology is a cohomology theory which performs the
reduction of the poisson algebra $\smooth{M}$ of smooth functions on a
poisson manifold $(M,P)$ by the ideal $I$ of functions which vanish on a
submanifold in the case where $I$ is a coisotropic ideal of $\smooth{M}$.
However, one might expect to able to perform the classical BRST construction
for the reduction of the poisson algebra $\smooth{M}$ by a coisotropic ideal
$I$ whether it is the ideal of functions vanishing on some submanifold or
not.  We will show that the reduction of a poisson manifold by an arbitrary
submanifold can always be thought of as the reduction of poisson algebras by
a suitable coisotropic ideal.

Following ${\Ref\Wilbour}$, let us restrict ourselves to certain interesting
ideals.
\Def{Let $J$ be an ideal in the poisson algebra $\P$.  $J$ is said to be an
{\sl associative ideal in} $\P$ if and only if
$$f^2\in J\ \implies\ f\in J.\ \eqn$$}

Notice that if $I$ is the ideal of functions which vanish on a submanifold
$M_o$ in a poisson algebra $\smooth{M}$ then $I$ is an associative ideal
since if $f(p)^2 = 0$ then $f(p) = 0$ for all points $p$ in $M_o$.
Associative ideals are interesting because of the following result from
${\Ref\Wilbour}$.

\Prop{Let $\P$ be a poisson algebra and $J$ an associative ideal.
Furthermore, let $J' = N(J)\intersect J$ where $N(J)$ is the normalizer of
$J$. If $N(J')$ is the normalizer of $J'$ then
$$ N(J) = N(J').\ \tag\ASSOCRESULT$$}

\Pf
Suppose that $f$ belongs to $N(J')$ for $J$ an associative ideal of
$\smooth{M}$ then for all $i$ in $J$, we have the equality
$$\comm{\comm{i^2}{f}}{f} = 2\comm{i}{f}^2 + 2i\comm{\comm{i}{f}}{f}.\
\tag\ASSOC$$
Since $i^2$ belongs to $J^2\subseteq J'$, $\comm{i^2}{f}$ belongs to $J'$
which in turn implies that $\comm{\comm{i^2}{f}}{f}$ belongs to $J'\subseteq
J$. In other words, the left hand side of equation $\ASSOC$ belongs to $J$.
Since in addition $2i\comm{\comm{i}{f}}{f}$ belongs to $J$, equation $\ASSOC$
implies that $\comm{i}{f}^2$ belongs to $J$ but since $J$ is an associative
ideal, this means that $\comm{i}{f}$ belongs to $J$.  In other words, $f$
belongs to $N(J)$ thereby proving that $N(J')\subseteq N(J)$.

Conversely, suppose that $f$ belongs to $N(J)$, then $\comm{f}{i'}$ belongs to
$J$ for all $i'$ in $J'$.  We need only show that $\comm{f}{i'}$ belongs to
$N(J')$ for all $i'$ in $J'$ to establish that $N(J)\subseteq N(J')$.  Notice
that for all $i$ in $J$ and $i'$ in $J$, we have $\comm{\comm{f}{i'}}{i} =
\comm{\comm{f}{i}}{i'} + \comm{f}{\comm{i}{i'}}$ but $\comm{f}{i}$ belongs to
$J$ and $\comm{i}{i'}$ belongs to $I$ therefore, we conclude that $f$ belongs
to $N(J)$.  This proves equation $\ASSOCRESULT$.
\qed

Therefore, if $\P$ is a poisson algebra and $I$ is an associative ideal, then
the reduction of $\P$ by $I$ is ${{N(I')}\over {I'}}$ which is naturally a
poisson algebra since $I'$ is a poisson ideal of $N(I')$.  In other words, we
have shown that the reduction of $\P$ by an associative ideal $I$ is the same
as the reduction of $\P$ by the ideal $I'$.

In the case where $(\Mtilde,\Ptilde)$ is the reduction of the poisson
manifold $(M,P)$ by $M_o$ and $I$ is the ideal of functions which vanish on
$M_o$ then we have the isomorphism of poisson algebras
$$\smooth{\Mtilde} \isomap {{N(I')}\over {I'}}.\ \tag\FINALISO$$

The usual classical BRST construction occurs when $(M,P)$ gives rise to a
symplectic manifold and $I$ is the ideal of functions which vanish on a
closed and embedded coisotropic submanifold of $M$.  In this case, $I' = I$
and $I$ is generated by a collection of so-called ``first-class constraints''
with respect to which the classical BRST complex is constructed.

The program that we wish to follow is now apparent.  We will replace the
role of $I$ by $I'$ in doing the classical BRST construction.  It differs
from the usual classical BRST construction since $I'$ is not generally the
ideal of functions which vanish on some submanifold of $M$.  Since $I'$ is a
coisotropic ideal in $\smooth{M}$, we expect many of the usual constructions
in classical BRST to generalize.  Although satisfactory from a purely
homological standpoint since the role of generators for $I$ is just
replaced by generators in $I',$ it may not be satisfactory in certain
physical applications. After all, in physical applications, we are usually
given constraints which generate $I$ and not $I'$ and, in general, there is
no natural way to get from a collection of constraints in $I$ to a collection
of constraints in $I'$.  Furthermore, it is often desirable to continue
working with the constraints in $I$ because these constraints might possess
some desirable covariance property that one is trying to preserve.  However,
things are not  quite as bad as they might seem.  For example, in the case
where $(M,P)$ gives rise to a symplectic manifold and $M_o$ is a symplectic
submanifold, then we have the isomorphism
$$\smooth{\Mtilde} \isomap {{N(I^2)}\over {I^2}}.\ \eqn$$
In this case, a collection of generators for $I$, say
$\Phi=(\phi_1,\ldots,\phi_k)$, naturally gives rise to a collection of
(first class) constraints in $I^2$ namely $\{\phi_i\phi_j\,|\,1\leq i\leq
j\leq k\}$.  These  generators for $I^2$ would preserve any ``Lorentz
covariance'' type properties of the original constraints.  However, we will
see that the constraints which generate $I^2$ will be infinitely
reducible.  We will be careful to take this into account.

\section{The Koszul-Tate Resolution Revisited}

The Koszul-Tate resolution is a complex which has nontrivial homology only in
zero dimension where it is isomorphic to $\smooth{M}/J$ where $J$ is any
ideal in $\smooth{M}$.  This complex performs the first step in
the reduction of $\smooth{M}$ by an ideal $J$ -- namely going from
$\smooth{M}$ to $\smooth{M}/J$. The Koszul-Tate resolution is a
generalization of the Koszul resolution due to Tate$^{\Ref\Tate}$ which is
performed by adjoining additional variables to the space of Koszul chains.
These additional variables will turn out to be the antighost sector of the
``ghosts for ghosts'' in the BRST complex.  Of particular interest is
the case where $J = I'$ for some ideal $I$ of functions which vanish on a
closed and embedded submanifold of a poisson manifold $(M,P)$.  In the next
section, we will see that when $(M,P)$ is a symplectic manifold and $I$ is
the ideal of functions which vanish on a closed and embedded symplectic
submanifold $M_o$ then $I' = I^2$ is always generated by infinitely
reducible constraints. In this section, we follow ${\Ref\FHST}$ in the
construction of a Koszul-Tate complex but for general ideals $J$ in
$\smooth{M}$ placing special emphasis upon the case of infinitely reducible
constraints.

Let us review the construction of the Koszul-Tate complex.  Let $J$ be an
ideal of $\smooth{M}$ generated by the elements $\Psi = (\psi_1, \psi_2,
\ldots, \psi_{m_o})$ (called {\sl constraints}). The usual Koszul
complex$^{\Ref\Lang}$  is constructed by introducing a free
$\smooth{M}$-module $V_1$ with a basis given by $\{\b{o}{a_o}\,|\,a_o =
1,\ldots,m_o\}$ whose elements are called  the {\sl antighosts of level} $0$.
In other words, $V_1$ consists of elements of the form $\sum_{a_o=1}^{m_o}
f^{a_o}\b{o}{a_o}$ where $f^{a_o}$ belongs to $\smooth{M}$ for all $a_o=1,
\ldots, m_o$.  The free module $V_1$ is given a subscript $1$ to indicate
that its elements are assigned a $\integ$-grading $1$ called the {\sl
antighost number}.  The space of Koszul chains is given by $\KT{o} =
\SS{V_1}$, the symmetric superalgebra over $V_1$.  In other words, $\KT{o}$
consists of all polynomials with coefficients in $\smooth{M}$ over the
antighosts where we regard these antighosts as being anticommuting
variables.   Another way to put it is that
$\KT{o}=\dsum_{b=0}^{m_o}\KT{o}_b$ forms a commutative and associative
superalgebra with unit graded by antighost number $b$ freely generated by the
antighosts. The Koszul differential is defined to be a $\Cinf(M)$-linear
graded derivation $\dk{o}:\KT{o}_{b+1}\to \KT{o}_{b}$ such that
$$\dk{o}\b{o}{a_o} = \psi_{a_o}.\ \tag\KOSZULDIFF$$
The homology of this complex, $H(\KT{o})$, is $\Cinf(M)/J$ in zero dimension.

We say that the constraints $\Psi$ are {\sl irreducible} if $\lambda^{a_o}$
belongs to $\smooth{M}$ and
$$ \sum_{a_o=1}^{m_o} \lambda^{a_o}\psi_{a_o} = 0\ \implies\
   \lambda^{a_o} \in J\ ,\ \forall a_o=1,\ldots, m_o.\ \tag\IRREDUCIBLE$$
If $\Psi$ are irreducible, then $\Psi$ forms a regular sequence in
$\smooth{M}$ and, therefore, the homology of the Koszul complex vanishes for
nonzero antighost number.  However, if $\Psi$ are not a set of
irreducible constraints, then we say that the constraints $\Psi$ are {\sl
reducible.} In this case, there will be nontrivial cycles at nonzero
antighost number.

If $\Psi$ are a collection of reducible constraints then there exists a
collection of functions on $M$, $\Z{1}{a_1}{a_o}$ (for $a_o=1,\ldots,m_o$ and
$a_1=1,\ldots,m_1$ for some $m_1$), which do not belong to $J\backslash 0$
such that
$$\sum_{a_o=1}^{m_o}\Z{1}{a_1}{a_o} \psi_{a_o} = 0\ \tag\REDUCE$$
for all $a_1$, and, for all functions $\lambda^{a_o}$, we have
$$\sum_{a_o=1}^{m_o}\lambda^{a_0}\psi_{a_0} = 0\ \implies\ \lambda^{a_0}\,=\,
\sum_{a_1=1}^{m_1} \Z{1}{a_1}{a_0} \rho^{a_1} \mod J\ \tag\COMPLETE $$
for some functions $\rho^{a_1}$.   If $\Z{1}{a_1}{a_o}$ exist which satisfy
equation $\REDUCE$ then $H_1(\KT{o})$ is nonzero since it contains
homologically nontrivial cycles of the form
$$\sum_{a_o=1}^{m_o}\Z{1}{a_1}{a_o}\b{o}{a_o}\ \tag\NONTRIVIAL$$
for all $a_1 = 1,\ldots,m_1$.  Furthermore, equation $\COMPLETE$ means that
the space of all nontrivial cycles in $\KT{o}_1$ are generated by such
elements.  We shall now utilize the method of Tate$^{\Ref\Tate}$ to remove
the nontrivial cycles in $\KT{o}_1$. Introduce a free $\smooth{M}$-module
$V_2$ which has a basis $\{\b{1}{a_1}\,|\,a_1 = 1,\ldots,m_1\}$.  The
elements in this basis are called antighosts of level $1$ and are assigned
antighost number $2$. A new space of chains is constructed from the old by
adjoining these new antighosts.  We define $\KT{1}=\SS{V_1\dsum V_2}$ where
$\SS{V_1\dsum V_2}$ is the symmetric superalgebra over the free
$\smooth{M}$-module $(V_1\dsum V_2)$, \ie\ $\KT{1}$ consists of all
polynomials with coefficients in $\smooth{M}$ over the commuting variables
$\b{1}{a_1}$ and the anticommuting variables $\b{o}{a_o}$.  This makes
$\KT{1} = \dsum_{b=0}^\infty\KT{1}_b$ into a commutative and associative
superalgebra with unit freely generated by the antighosts graded by antighost
number. The new differential, $\dk{1}:\KT{1}_{b+1}\to \KT{1}_{b}$ is a
$\Cinf(M)$-linear graded derivation which extends $\dk{o}$ by mapping
$\b{o}{a_o}$ into the nontrivial cycles in equation $\NONTRIVIAL$, \ie\
$$\dk{1}\b{1}{a_1} = \sum_{a_o=1}^{m_o}\Z{1}{a_1}{a_o}\b{o}{a_o}.\
\tag\KOSZULZERO$$
Equation $\REDUCE$ insures that $\dk{1}^2 = 0$.  These new antighosts kill
off the nontrivial cycles in $\KT{o}_1$ while leaving the homology in zero
dimension unchanged, \ie\ $H_o(\KT{1}) = \Cinf(M)/J$ and $H_1(\KT{1})=0$.
However, there may be nontrivial cycles in $\KT{1}_2$ either because they
were there to begin with in the Koszul complex or because we have introduced
them by choosing overcomplete $\Z{1}{a_1}{a_o}$.

This procedure can be carried out for higher antighost numbers.$^{\Ref\FHST}$
Suppose that there exists a collection of functions $\Z{i}{a_i}{a_{i-1}}$
which do not belong to $J\backslash 0$ where $i = 1, \ldots, L$ and $a_i=
1,\ldots,m_i$ that satisfy equations $\REDUCE$ and $\COMPLETE$ as well as the
equations
$$\sum_{a_{j-1}=1}^{m_{j-1}}\Z{j}{a_j}{a_{j-1}}\Z{j-1}{a_{j-1}}{a_{j-2}} = 0
\mod J\ \eqn$$
and
$$\sum_{a_{j-1}=1}^{m_{j-1}}\lambda^{a_{j-1}}\Z{j-1}{a_{j-1}}{a_{j-2}}=0\mod
J\ \implies \  \lambda^{a_{j-1}} = \sum_{a_j=1}^{m_j} \rho^{a_j}
\Z{j}{a_j}{a_{j-1}}\mod J\ \eqn$$
all $j=2,\ldots,L$ and $a_j=1,\ldots,m_j$.  The number $L$ is called the
{\it order of reducibility} of this system. It is defined to be the last $i$
for which $\Z{i}{a_i}{a_{i-1}}$ is nonzero.  It is possible for $L$ to be
infinite but let us assume that $L$ is finite for now.

Suppose that we have constructed the Koszul-Tate resolution up to level $i$
where $i < L$.  That is, for all $j=0,\ldots,i$, we have introduced free
$\smooth{M}$-modules $V_{j+1}$ which are assigned antighost number $j+1$ and
are spanned by the antighosts of level $j$,  $\{\b{j}{a_{j}}\,|\, a_{j} =
1,\ldots,m_i\}$, such that the space of chains is $\KT{i}=\SS{\V{i}}$
where $\V{i} = \Dsum_{j=1}^{i+1} V_j$.  Furthermore, we have defined the
differential $\dk{i}:\KT{i}_{b+1}\,\to\,\KT{i}_b$ by equation $\KOSZULZERO$
and
$$ \dk{i}\b{j}{a_j} = \sum_{a_{j-1}=1}^{m_{j-1}}\Z{j}{a_j}{a_{j-1}}
\b{j-1}{a_{j-1}} + \M{j}{a_j}\  \eqn$$
for all $j=1, \ldots, i$.  The homology of the complex $\dk{i}:\KT{i}_{b+1}
\,\to\, \KT{i}_b$ vanishes for antighost number $b=1, \ldots, i$ since, by
construction, we have removed all of the nontrivial cycles up to antighost
number $i$. The nontrivial cycles in $\KT{i}_{i+1}$ are generated by
$$\sum_{a_i=1}^{m_i}\Z{i+1}{a_{i+1}}{a_i}\b{i}{a_i}+\M{i+1}{a_{i+1}}\
\tag\GENCYCLES$$
where $\M{i}{a_i}$ has antighost number $i+1$ and contains only antighosts of
level less than $i$.  As before, $\M{i+1}{a_{i+1}}$ is arbitrary up to a
boundary.  We introduce a free $\smooth{M}$-module $V_{i+1}$ which
has antighost number $i+2$ and a basis $\{\b{i+1}{1}, \b{i+1}{2}, \ldots,
\b{i+1}{m_{i+1}}\}$. Let us define the level $i+1$ Koszul-Tate chains
by $\KT{i+1}=\SS{\V{i+1}}$.  The differential on this complex is a
$\smooth{M}$-linear graded derivation $\dk{i+1}:\KT{i+1}_{b+1}\to\KT{i+1}_b$
which extends $\dk{i}$ via
$$\dk{i+1}\b{i+1}{a_{i+1}} = \sum_{a_i=1}^{m_i}\Z{i+1}{a_{i+1}}{a_i}
\b{i}{a_i} + \Mp{i+1}{a_{i+1}}\ \tag\TRIALDEF$$
for some $\Mp{i+1}{a_{i+1}}$ consisting of antighosts with level less than
$i$ such that $\dk{i+1}^2\b{i+1}{a_{i+1}} = 0$.  It is easy to verify that
any such differential must have $\Mp{i+1}{a_{i+1}} = \M{i+1}{a_{i+1}}$ up to
a $\dk{i}$ boundary$^{\Ref\FHST}$ and, therefore, its homology satisfies
$$H_o(\KT{i+1}) = \Cinf(M)/J\  {\rm  and}\  H_b(\KT{i+1}) = 0,\ \forall b=
1,\ldots, i+1.\ \tag\ACYCLICGEN$$
In other words, any differential $\dk{i+1}$ which satisfies $\TRIALDEF$ and
agrees when acting upon lower antighosts with $\dk{i}$, is a bona fide
Koszul-Tate differential, \ie\  its associated homology satisfies
$\ACYCLICGEN$. We will use this result later when constructing the BRST
charge.

This construction proceeds until $i$ reaches the level $L$.  At this point,
the complex $\KT{L}=\SS{\V{L}}$ with differential
$\dk{L}:\KT{L}_{b+1}\to\KT{L}_b$ forms an acyclic resolution of $\Cinf(M)/J$.
$^{\Ref\Tate}$

In the case where the system is infinitely reducible, this procedure is
iterated an infinite number of times introducing a infinite level of
antighosts.  This gives rise to the space of Koszul-Tate chains $\KT{\infty}
= \SS{\V{\infty}}$ and the differential
$\dk{\infty}:\KT{\infty}_{b+1}\to\KT{\infty}_b$ is a graded derivation which
acts upon each antighost via equation $\TRIALDEF$.  The Koszul-Tate complex
still forms an acyclic resolution of $\Cinf(M)/J$ since all of the nontrivial
cycles at any given antighost number have been removed by the same
construction.

\section{Infinite Reducibility}

Let $i:M_o\into M$ be a closed and embedded submanifold of codimension $k$
where $k\geq 2$ and $I$ be the ideal of functions in $\Cinf(M)$ which vanish
on $M_o$.  In this section, we present a particular collection of elements
which generate $I^2$ (essentially the collection of products of constraints)
that are necessarily infinitely reducible.  This is of particular interest in
the case where $M_o$ is a symplectic submanifold of $M$, \eg\ it arises as the
zero locus of a collection of so-called ``second class constraints''. In this
case, we recall that $I' = I^2$ is a coisotropic  ideal of $\smooth{M}$ --
the ideal with respect to which we would construct the Koszul-Tate complex
and, eventually, the BRST complex.

Let us begin with a special case. Let $i:M_o\into M$ be a closed and embedded
submanifold of codimension $2$ and $I$ be the ideal of functions which vanish
on $M_o$. Let us assume that $I$ is generated by the irreducible constraints
$\Phi=(\phi_1,\phi_2)$, \ie\ $\Phi$ are irreducible constraints
and
$$I = \{\phi_1 f + \phi_2 g\,|\, f,g\in\Cinf(M)\}.\ \eqn$$
In this case, the ideal $I^2$ is generated by the elements $\Psi\equiv
(\psi_1,\psi_2,\psi_3) \equiv (\phi_1^2,\phi_1\phi_2,\phi_2^2)$, \ie\
$$I^2 = \{\sum_{a_o=1}^{3}\alpha^{a_o}\psi_{a_o}\,|\,\alpha^{a_o} \in
\Cinf(M)\}\,.\ \eqn$$
We will show that $\Psi$ are necessarily infinitely reducible.

Suppose that
$$\lambda^{1} \phi_1^2 + \lambda^{2} \phi_1 \phi_2 + \lambda^{3} \phi_2^2 = 0\
\tag\REDEX$$
for some functions $\lambda^{a_o}$ then $\lambda^1\phi_1^2$ belongs to the
ideal generated by $\phi_2$.  We would like to conclude that $\lambda^{1} =
\rho^{1} \phi_2$ for some function $\rho^{1}$.   As usual, this is done first
locally and then extended globally.

\Def{Let $i:M_o\,\to\,M$ be a closed and embedded codimension $k$ submanifold
of an $n$-dimensional manifold $M$ such that $M_o$ is the zero locus of a map
$\Phi = (\phi_1,\ldots,\phi_k):M\,\to\,\reals^k$.  A {\sl regular coordinate
chart $x:U\,\to\,\reals^n$ of $M_o$ in $M$ with respect to $\Phi$} is a
coordinate chart $x:U\,\to\,\reals^n$ such that $U\intersect M_o \not=
\emptyset$, $x = (\phi_1,\ldots,\phi_k, y_1, \ldots, y_{n-k})$, and $y =
(y_1,\ldots, y_{n-k}):M_o\intersect U\,\to\,\reals^{n-k}$ is a coordinate
chart for $U\intersect M_o$.}

Let $\U = \{U_\alpha, M\backslash M_o\}$ be a cover of $M$ where
$\{x_\alpha:U_\alpha\,\to\,\reals^n\}$ is a collection of regular coordinate
charts of $M_o$ in $M$ with respect to $\Phi$ and let $\{\sigma_\alpha,
\sigma^\prime\}$ be a partition of unity subordinate to this cover.  In the
regular neighborhood $U_\alpha$, the fact that $\lambda^1\phi_1^2$ belongs to
the ideal generated by $\phi_2$ implies that  $\lambda^1 = h_\alpha\phi_2$
for some $h_\alpha$ in $\Cinf(U_\alpha)$. This is the case since the $\phi_i$
are $2$ of the coordinates of $U_\alpha$. We can extend $h_\alpha$ to a
function on $M$ by using partitions of unity, \ie\ $\rho^1 \equiv \sum_\alpha
h_\alpha\sigma_\alpha + h^\prime\sigma^\prime$ where $h^\prime$ is any
function on $M\backslash M_o$ so that $\lambda^1 = \rho^1 \phi_2$ globally.
Actually, this result is nothing more than the fact that $\Phi$ forms a
regular sequence in $\Cinf(M)$.

Similarly, $\REDEX$ tells us that $\lambda^3\phi_2^2$ belongs to the ideal
generated by $\phi_1$ and, therefore, $\lambda^{2} = \rho^{2} \phi_1$ for
some function $\rho^{2}$. Plugging this into  $\REDEX$, we obtain $\phi_1
\phi_2(\rho^{2} \phi_1 + \lambda^{2} + \rho^{1} \phi_2) = 0.$ Working in a
regular cover and globalizing the result, we conclude that  $\lambda^{2} =
-\rho^{2} \phi_1 - \rho^{1} \phi_2$ and we define
$$\eqnalign{\Z{1}{1}{1} = \phi_2,\ \Z{1}{1}{2} &= -\phi_1,\ \Z{1}{1}{3} = 0\cr
\Z{1}{2}{1} = 0, \ \Z{1}{2}{2} &= -\phi_2,\ \Z{1}{2}{3} = \phi_1.\ \eqn\cr}$$
which satisfies $\lambda^{a_o} = \sum_{a_1=1}^3\Z{1}{a_1}{a_o}\rho^{a_1}$ for
all $a_o= 1,2,3$.  Since $\Z{1}{a_1}{a_o}$ do not belong to $I^2\backslash 0$
and they satisfy $\REDUCE$ and $\COMPLETE$, this concludes the analysis of
reducibility at level one.

What of the second level?  Suppose that $\sum_{a_1=1}^3\lambda^{a_1}
\Z{1}{a_1}{a_o}$ belongs to $I^2\backslash 0$ for all $a_o$ for some functions
$\lambda^{a_1}$. Plugging in $a_o = 1$, we see that $\lambda^1 \phi_2$
belongs to $I^2$ and, therefore, $\lambda^1$ must belong to $I$. Similiarly,
plugging in $a_o = 3$  implies that $\lambda^2$ belongs to $I$. Finally, if
$\lambda^1$ and $\lambda^2$ both belong to $I$ then the equation which
results from setting $a_o = 2$ is automatically satisfied.  In other words,
there exists a collection of functions $\rho^{a_2}$ where $a_2 = 1,\ldots, 4$
such that
$$\lambda^1 = \rho^1 \phi_1 + \rho^2 \phi_2\ {\rm and}\ \lambda^2 = \rho^3
\phi_1 + \rho^4 \phi_2.\ \tag\LAMBtwo$$
Let us now define a collection of functions $\Z{2}{a_2}{a_1}$ where $a_2 =
1, \ldots, 4$ which do not belong to $I^2\backslash 0$ (but belong to $I$)
via
$$\eqnalign{\Z{2}{1}{1} &= \Z{2}{3}{2} = \phi_1\cr
            \Z{2}{2}{1} &= \Z{2}{4}{2} =\phi_2\ \eqn\cr}$$
where all others vanish.  Equation $\LAMBtwo$ is then $\lambda^{a_1} =
\Z{2}{a_2}{a_1}\rho^{a_2}$ and this completes the analysis of reducibility at
level two.

A similar computation for level three gives us the $\Z{3}{a_3}{a_2}$ where
$a_3 = 1, \ldots, 8$ are given by
$$\eqnalign{\Z{3}{1}{1} = \Z{3}{3}{2} = \Z{3}{5}{3} =& \Z{3}{7}{4} =
\phi_1,\cr \Z{3}{2}{1} = \Z{3}{4}{2} = \Z{3}{6}{3} =& \Z{3}{8}{4} = \phi_2,\
\eqn\cr}$$
and all other $\Z{3}{a_3}{a_2}$ vanish.  This pattern continues for an
infinite number of levels.  The $\Z{i}{a_i}{a_{i-1}}$ (for $i\geq 2$) will be
a collections of functions  where $a_i = 1,\ldots, 2^i$ and $a_{i-1} =
1,\ldots 2^{i-1}$ which belong to $I$ but not to $I^2\backslash 0$ given by
$$\eqnalign{\Z{i}{1}{1} = \Z{i}{3}{2} =, \cdots, = \Z{i}{2^i-1}{2^i-1} =&
\phi_1,\cr \Z{i}{2}{1} = \Z{i}{4}{2} =, \cdots, = \Z{i}{2^i}{2^i-1} =&
\phi_2,\ \eqn\cr}$$
and all other $\Z{i}{a_i}{a_{i-1}}$ vanish.  This concludes our proof of the
infinite reducibility of $\Psi$.

Let us now consider the more general case where $i:M_o\into M$ is a closed
and embedded submanifold with codimension $k$ where $k > 2$ but where $I$ is
still generated by a collection of irreducible constraints
$\Phi=(\phi_1,\ldots,\phi_k)$.  In this case, $I^2$ will be generated by the
collection of elements $\Psi = \{\phi_i\phi_j\,|\,i\geq j = 1,\ldots, k\}$.
Since $\Psi$ contains $\{\phi_1^2,\phi_1\phi_2, \phi_2^2\}$ as a subset and
we have shown that subset is infinitely reducible, $\Psi$ is itself
infinitely reducible. After all, the introduction of the additional
generators does not remove the reducibility of the original set of
generators.

Let us now relax the assumption that $I$ is generated by irreducible
constraints. If $I$ is generated by the reducible constraints $\Phi =
(\phi_1,\phi_2,\ldots, \phi_l)$ for some $l>k$, then $I^2$ will still be
generated  by the elements $\Psi = \{\phi_i\phi_j\, | \, i\leq j =
1,\ldots,l\}$.   The fact that $\Phi$ are reducible will only mean that there
are more relations between the various elements in $\phi_i\phi_j$, not less.
Therefore, $\Psi$ will still be an infinitely reducible set of
constraints. Another way to see this is that about every point in $M_o$,
there exists an open neighborhood in $M$ containing it, $U$, such that a
subset of $k$ elements in $\Phi$ are regular constraints in $\Cinf(U)$.
These regular constraints are locally infinitely reducible following the
argument given above.  Suppose it were true that these constraints are
globally finitely reducible then this would imply that the constraints would
locally be finitely reducible which would be a contradiction.

We have just shown the following theorem:

\Thmtag\InfiniteReducible{Let $i:M_o\into M$ be a closed and embedded
submanifold of codimension $k\geq 2$.  If $I$ is the ideal of functions which
vanish on $M_o$ generated by elements $(\phi_1,\phi_2,\ldots,\phi_k)$ then
the collection of elements which generate $I^2$ given by
$\{\phi_i\phi_j\,|\,i\leq j=1,\ldots,k\}$ are necessarily infinitely
reducible.}

\par

\section{The BRST Complex}

In this section, we construct the BRST complex by extending the Koszul-Tate
complex through the introduction of ghosts. We show that the space of BRST
cochains forms a poisson superalgebra which is graded by an integer called
the ghost number.  The BRST differential is exhibited as an inner derivation
by an element with ghost number $1$ called the BRST charge. Therefore, the
associated cohomology inherits the structure of a poisson superalgebra
graded by ghost number.  If $J$ is a coisotropic ideal of $\smooth{M}$ then
we will see that BRST cohomology is isomorphic as poisson algebras in zero
dimension to $N(J)/J$.

Let $J$ be generated by elements $\psi_{a_o}$ where $a_o = 1,\ldots,m_o$.
Furthermore, let us assume that  this system has order of reducibility $L$.
Suppose that the Koszul-Tate complex has been constructed up to level $i <
L$, \ie\ we have introduced the free $\smooth{M}$-module graded by antighost
number $\V{i} = \Dsum_{j=1}^{i+1} V_j$ where $V_{j+1}$ is  spanned by
$\{\b{j}{a_j}\,|\,a_j= 1,\dots,m_j\}$, the antighosts at level $j$.  Let the
dual free $\smooth{M}$-module of $\V{i}$ be denoted by $\V{i}^* =
\Dsum_{j=1}^{i+1} V_j^*$ where each $V_{j+1}^*$ has a basis dual to the
antighosts at level $j$ called the {\it ghosts} at level $j$ which will
be denoted by $\{\c{j}{a_j}\,|\,a_j = 1,\ldots,m_j\}$.  In other words, if
$\langle\cdot\,,\,\cdot\rangle:\V{i}^* \dsum \V{i} \to \smooth{M}$ is the
dual pairing then
$$\langle \c{j}{a_j}\,,\,\b{l}{d_l}\rangle\,=\,\delta_j^l\delta_{a_j}^{d_l}\
\eqn$$
for all $j, l = 0, \ldots, i$, $a_j = 1,\ldots,m_j$, and $d_l=1,\ldots,m_l$.
These ghosts have their own $\integ$-grading  called the {\sl $c$-number.}
The $c$-number of $\c{j}{a_j}$ is defined to be $j+1$ and the $c$-number of
$\b{j}{a_j}$ is defined to be $0$. This is similar to the antighost number,
which we will now call the {\sl $b$-number}, since the $b$-number of
$\b{j}{a_j}$ is $j+1$ and we will define the $b$-number of $\c{j}{a_j}$ to be
zero.  Finally, the {\sl ghost number}  is a $\integ$-grading defined to be
the difference of the $c$-number and the $b$-number.  We will often denote the
ghost number of an object $u$ by $\abs{u}$.

The grading by ghost number makes $\V{i}\dsum\V{i}^*$ into a $\integ$-graded
free $\smooth{M}$-module.  Consider $\SS{\V{i}\dsum\V{i}^*}$, the associative
and commutative superalgebra (graded by ghost number) freely generated by the
ghosts and antighosts over $\smooth{M}$.  It can be endowed with the
structure of a poisson superalgebra where the poisson bracket satisfies
$\comm{a}{b} = -(-)^{\abs{a}\abs{b}}\comm{b}{a}$ for all $a$ and $b$ in
$\symmetric(\V{i}\dsum\V{i}^*)$ with definite ghost number.  This poisson
bracket extends the poisson bracket on $\smooth{M}$ via the dual pairing, \ie
$$\comm{\c{l}{a_l}}{\c{j}{d_j}} = \comm{\b{l}{a_l}}{\b{j}{d_j}}
      = \comm{\c{l}{a_l}}{f} = \comm{\b{l}{a_l}}{f} = 0,\ \tag\PBone$$
$$\comm{\c{l}{a_l}}{\b{j}{d_j}} = \delta^l_j\delta^{a_l}_{d_j}\ \tag\PBtwo$$
for all $j, l = 0, \ldots, i$, $a_j=1, \ldots, m_j$, $d_l = 1, \ldots, m_l$,
and $f$ in $\smooth{M}$ as well as the intertwining relation
$$\comm{a}{b c} = \comm{a}{b} c + (-1)^{\abs{a}\abs{b}} b\comm{a}{c}\
\tag\PBthree$$
for all $a,b,c$ in $\SS{\V{i}\dsum\V{i}^*}$ with definite ghost
numbers.    This bracket preserves the ghost number grading but does not
preserve the $(c,b)$-number bigrading.  Therefore, $\SS{\V{i}\dsum\V{i}^*}$
is a poisson superalgebra graded by ghost number.

Let us think of $\SS{\V{i}\dsum\V{i}^*}$ as the space of polynomials with
coefficients in $\smooth{M}$ in the $\integ_2$-graded variables
$\{\b{j}{a_j}, \c{j}{a_j}\,|\, j=1, \ldots, i; \ a_j = 1, \ldots, m_j \}$.
The space of BRST cochains of reducibility level $i$ with ghost number
$g$, $\K{i}^g$, consists of the space of all {\sl infinite formal sums} with
coefficients in $\smooth{M}$ over the $\integ_2$-graded variables
$\{\b{j}{a_j}, \c{j}{a_j}\,|\, j=1, \ldots, i; \ a_j = 1, \ldots, m_j \}$
which have ghost number $g$.  The BRST cochains at various ghost numbers for
a certain reducibility level $i$ assemble into $\K{i} = \Dsum_{g\in\integ}
\K{i}^g$, \ie\  finite sums of elements at different ghost numbers.  $\K{i}$
is endowed with the structure of a poisson superalgebra which naturally
extends the one defined above for $\SS{\V{i}\dsum\V{i}^*}$. $\K{i}$ forms a
poisson superalgebra graded by ghost number but the poisson
bracket does not preserve the $(c,b)$-bigrading. For a system of reducibility
level $L$, the total space of BRST cochains is given by the poisson
superalgebra $\K{L}$ graded by ghost number.

The reason that we allow certain infinite formal sums in the space of
BRST cochains is so that the BRST differential can be given as an inner
derivation by an element $Q$ with ghost number $1$. In the case of
infinite reducibility, $Q$ must be the sum of  an infinite number of monomials.
Furthermore, imposing that $\K{i}$ consists of only finite sums of elements
with different ghost number insures that in the infinitely reducible case,
$\K{\infty}$ has a well-defined poisson bracket.  For example, consider the
poisson bracket of infinite sums of elements at different ghost number, \eg
$$ \sum_{j=0}^\infty\sum_{a_j=1}^{m_j} \comm{f^{(j,a_j)}\b{j}{a_j}}
{g_{(l,d_l)}\c{l}{d_l}} = \sum_{j=0}^\infty\sum_{a_j=1}^{m_j}
(f^{(j,a_j)}\, g_{(j,a_j)} + \comm{f^{(j,a_j)}}{g_{(l,d_l)}}\,\b{j}{a_j}\,
\c{l}{d_l})\,.\  \tag\NotPB$$
The first term on the right hand side need not converge.  Therefore,
$\K{\infty}$ would not form a poisson algebra if all infinite formal sums of
elements at different ghost numbers were allowed. This problem does not arise
if we define $\K{\infty}$ to consists of only finite sums of elements at
different ghost numbers.   Of course, it may be possible that $\K{\infty}$
could be allowed to contain some subset of infinite sums of elements at
different ghost numbers.  After all, the first term on the right hand side of
$\NotPB$ will converge for certain choices of functions $f^{(j,a_j)}$ and
$g_{(l,d_o)}$, \eg\ it may be possible to redefine the space of BRST cochains
to be the completion of $\SS{\V{i}\dsum\V{i}^*}$  with respect to some norm.
This resulting space, if it can be shown to respect the constructions of this
paper, would then be a possible space of BRST cochains.   However, for our
purposes, it is sufficient to consider the case where the space of BRST
cochains is a finite sum of elements at different ghost numbers.

Let $\K{i}^{c,b}$ be the set of all elements in $\K{i}^{c-b}$ which have the
correct $(c,b)$-number bigrading. In fact, we can think of elements in
$\K{i}^g$ as infinite formal sums of elements in $\K{i}^{c,b}$ where $c-b =
g$.  As stated before, $\K{i}^g$ forms a poisson algebra graded by ghost
number but the $(c,b)$-number bigrading of $\K{i}$ is not respected by the
poisson bracket.  Nonetheless, this bigrading does provide an additional
structure. We can define a filtration of $\K{i}$ by $F^p\K{i} = \Dsum_{c\geq
p, b\geq 0}\K{i}^{c,b}$ so that
$$ \K{i}\, =\,F^o\K{i}\,\supseteq\,F^1\K{i}\,\supseteq\,F^2\K{i}\,
\supseteq\,F^3\K{i}\,\supseteq\,\cdots.\ \tag\FILTRATION $$
Since our poisson bracket satisfies $\comm{F^p\K{i}}{F^q\K{i}}\subseteq
F^{p+q} \K{i}$ and similarly for the associative multiplication, $\K{i}$ is a
{\sl filtered poisson superalgebra}.  Let us denote the space of elements in
$F^p\K{i}$ with ghost number $g$ by $F^p\K{i}^g$.

Suppose that there exists a sequence of maps
$$\cdots\,\mapright{\D{i}}\,\K{i}^{g-1}\,\mapright{\D{i}}\,\K{i}^g\,
\mapright{\D{i}}\,\K{i}^{g+1}\,\mapright{\D{i}}\,\cdots.\ \eqn$$
These maps naturally break up under the filtration degree into
$\D{i} = \d{i}{o} + \d{i}{1} + \d{i}{2} + \cdots$ where
$\d{i}{j}:\K{i}^{c,b}\to\K{i}^{c+j,b+j-1}$ for all $j\geq 0$.
We will find this decomposition to be useful in the next section.

It remains to introduce the BRST differential $D:\K{L}^g \to \K{L}^{g+1}$
which is a certain poisson derivation, \ie\  $D$ should satisfy
$$D\comm{u}{v} = \comm{Du}{v} + (-1)^{\abs{u}}\comm{u}{Dv}\ \eqn$$
where $u$ and $v$ are elements in $\K{L}$ with definite ghost number.
The reason that this property is desirable is that such a differential
insures that the BRST cohomology $H_D$ forms a poisson superalgebra graded by
ghost number.  It can be shown $^{\Ref\DuVi}$ that any poisson derivation on
$\K{L}$ which increases ghost number by $1$ is an inner poisson derivation,
\ie\  there exists an element $Q\in\K{L}^1$ such that $D = \comm{Q}{\cdot}$.
The fact that $D^2 = 0$ is equivalent to asserting that $\comm{Q}{Q}$ lies in
the center of $\K{L}$. The BRST charge is a particular element $Q$ in
$\K{L}^1$ which satisfies $\comm{Q}{Q} = 0$ such that its associated
cohomology in zero dimension is isomorphic as poisson superalgebras to
$N(J)/J$.  We will construct $Q$ in the next section.

\section{Construction of the BRST Charge}

In this section, we construct the BRST charge using a refinement of the
methods in ${\Ref\Stasheff}$ and ${\Ref\TKJMFReview}$. It is a BRST cochain
$Q$ with ghost number $1$ which satisfies
$$\comm{Q}{Q} = 0.\ \tag\QQ$$
The BRST differential $\D{L}:\K{L}^g\,\to\,\K{L}^{g+1}$ is given by the inner
derivation by this element
$$ \D{L} = \comm{Q}{\cdot}.\ \eqn$$
Notice that $\D{L}^2 = 0$ because of equation $\QQ$ and the Jacobi identity.
The BRST differential $\D{L}$ will often be denoted by $D$ for short.

The main result of this section can be summarized by the following.

\Thm{Let $J$ be a coisotropic ideal of $\smooth{M}$,
$\Psi=(\psi_1,\ldots,\psi_{m_o})$ be elements which generate $J$, and $\K{L}$
be the space of BRST cochains with respect to these constraints.  There
exists an element $Q$ in $\K{L}^1$ satisfying equation $\QQ$ such that
$$ Q = \sum_{a_o=1}^{m_o} \c{o}{a_o}\psi_{a_o} +
\sum_{i=1}^{L}\sum_{a_i=1}^{m_i}\sum_{a_{i-1}=1}^{m_{i-1}}\c{i}{a_i}
\Z{i}{a_i}{a_{i-1}}\b{i-1}{a_{i-1}} + {\rm etc}\ \tag\QBoundaryCond$$
where {\rm etc} consists of terms with at least two ghosts and one antighost
or terms with at least two antighosts and one ghost.  Furthermore, we can
replace $Q\mapsto Q+\d{L}{o}\lambda$ for any $\lambda$ in $\K{L}^2$
and still satisfy equations $\QQ$ and $\QBoundaryCond$ where $\d{L}{o}$ is
the Koszul-Tate differential.}

Let us begin by observing that that the filtration of $\K{i}$ defined in
equation $\FILTRATION$ is unbounded, in general.  Any element $x$ in
$\K{i}^g$ can be written as the sum $ x=\sum_{j=0}^\infty x_j$ where $x_j$
has $c$-number $j$.  In particular, we can decompose the BRST charge, if it
exists, into the (possibly infinite) sum
$$Q = Q_o+Q_1+Q_2+\cdots.\ \eqn$$
We will construct $Q$ inductively by constructing the $Q_{i+1}$ from
$Q_o,\ldots,Q_i$.  Let us begin with the definition
$$ Q_o = \sum_{a_o=1}^{m_o}\c{o}{a_o}\psi_{a_o}\ \eqn$$
then we see that $Q_o^2$ belongs to $F^2\K{0}^2$ since
$$Q_o^2 = \sum_{j_o,k_o,l_o=1}^{m_o} \stc{l_o}{j_o}{k_o}\c{o}{\,j_o}
\c{o}{k_o}\psi_{l_o}\ \eqn$$
where the structure functions $\stc{l_o}{j_o}{k_o}$ are defined by
$$\comm{\psi_{j_o}}{\psi_{k_o}} = \sum_{l_o=1}^{m_o}\stc{l_o}{j_o}{k_o}\,
\psi_{l_o}\ \eqn$$
for all $j_o, k_o = 1, \ldots, m_o$.

The right hand side of this equation looks like the image under the Koszul
differential of the $Q_1$-term in the BRST charge for irreducible constraints.
This observation forms the basis for the inductive construction of $Q$ that
follows.

Suppose that $Q_o$ is defined as above and there exists $ Q_j
\in\K{j}^{j+1,j}$ for all $j=1,\ldots, \r_i$ where $r_i = {\rm min}(i,L)$
such that
$$ Q_j = \sum_{a_j=1}^{m_j}\c{j}{a_j}(\sum_{a_{j-1}=1}^{m_{j-1}}
\Z{j}{a_j}{a_{j-1}}\b{j-1}{a_{j-1}}+ \M{j}{a_j}) + N_j\ \tag\INDONE $$
such that $N_j\in\K{j-1}$ contains at least two $c$'s and  the $\M{j}{a_j}$
are the same ones which appear in equation $\GENCYCLES$ and which contain at
least two $b$'s.  Furthermore, let us suppose that for all $j=0,\ldots,i$, we
have
$$ \comm{R_j}{R_j} \in F^{j+2}\K{j}^2\ \tag\INDTWO$$
where
$$ R_j = \sum_{l=0}^j Q_l.\ \eqn$$
Define $\D{i}:\K{\r_i}^g\,\to\,\K{\r_i}^{g+1}$ by $ \D{i} =
\comm{R_i}{\cdot}$ and decompose it by $c$-number, \ie
$$\D{i} = \sum_{j=0}^\infty\d{i}{j}\ \tag\BRSTDSUM$$
where $\d{i}{j}:\K{\r_i}^{c,b}\,\to\,\K{\r_i}^{c+j,b+j-1}.$
In particular, we have
$$\eqnalign{\d{i}{o} &= \sum_{a_o}^{m_o}\comm{\c{o}{a_o}}{\cdot}\psi_{a_o}+\cr
&\sum_{j=1}^{\r_i}\,\sum_{a_j=1}^{m_j}\,\comm{\c{j}{a_j}}{\cdot}
(\sum_{a_{j-1}=1}^{m_{j-1}}\Z{j}{a_j}{a_{j-1}} \b{j-1}{a_{j-1}}+ \M{j}{a_j})
(-1)^{(j+1)\abs{\cdot}}.\  \eqn}$$
This implies that
$$\d{i}{o}\c{l}{a_l} = 0\qquad {\rm and}\qquad \d{i}{o}\b{o}{a_o} =
\psi_{a_o}\ \eqn$$
for all $a_l=1,\ldots,m_l$; $l=0,\ldots,\r_i$; and
$$\d{i}{o}\b{j}{a_j} = \sum_{a_{j-1}=1}^{m_{j-1}} \Z{j}{a_j}{a_{j-1}}
\b{j-1}{a_{j-1}} + \M{j}{a_j},\ \eqn$$
for all $a_j= 1,\ldots,m_j$ and $j=1,\ldots,\r_i$.  It is just the level
$\r_i$ Koszul-Tate differential acting upon $\K{r}$ extended to act trivially
upon the ghosts.  From the construction of the Koszul complex, we have a
differential complex for each $c$-number and for all values of $j=0,\ldots,
L$ given by
$$\cdots\,\mapright{\d{j}{o}}\,\K{j}^{c,b+1}\,\mapright{\d{j}{o}}\,
\K{j}^{c,b}\, \mapright{\d{j}{o}}\,\K{j}^{c,b-1}\,\mapright{\d{j}{o}}
\,\cdots\ \eqn$$
whose homology $H(\K{j}^{c,\cdot},\d{j}{o})$ satisfies
$$H_b(\K{j}^{c,\cdot},\d{j}{o}) = \cases{0 & for $b=1,\ldots,j$\cr
                                      \K{j}^{c,0}& for $b=0$\cr}.\
\tag\EXTENDEDKT$$
If $j=L$ then we have the acyclicity condition
$$H_b(\K{L}^{c,\cdot},\d{L}{o}) = \cases{0 & for $b>0$\cr
                                      \K{L}^{c,0}& for $b=0$\cr}.\
\tag\FULLEXTENDEDKT$$
In order to avoid annoying factors of two which will otherwise arise, define
for elements $O$ in $\K{L}$ with odd ghost number
$$ O^2 = \half\, \comm{O}{O}.\ \eqn$$
then $\D{i}R_i^2 = \comm{R_i}{R_i^2} = 0$.  However, equation $\BRSTDSUM$
implies that
$$\d{i}{o} R_i^2 = -\d{i}{1} R_i^2 - \d{i}{2} R_i^2 -\d{i}{2} R_i^2-\cdots\,.\
\tag\REQ$$
Since $R_i^2$ is in $F^{i+2}\K{i}^2$ the filtration degrees of the $\d{i}{j}$
which appear on the right hand side of the equation $\REQ$, imply that
$$ \d{i}{o} R_i^2 \in F^{i+3}\K{i}^3.\ \eqn$$
Let $X_i\in\K{i}^{i+2,i}$ be the piece of $R_i^2$ which belongs to
$\K{i}^{i+2,i}$ then the previous equation implies that
$$\d{i}{o} X_i = 0.\ \eqn$$

Since $H_i(\K{r}^{i+2,\cdot},\d{i}{o}) = 0$, there exists an element $Y_i$ in
$\K{i}^{i+2,i+1}$ such that
$$X_i = -\d{i}{o} Y_i\,.\ \tag\YEQ$$
Of course, we notice that the $Y_i$ which we have chosen is hardly
unique.  In fact, equation $\YEQ$ is invariant under the shift
$Y_i\,\mapsto\,Y_i+U_i$ where $U_i$ is a  $\d{i}{o}$ closed cycle in
$\K{\r_i}^{i+2,i+1}$.

There are two cases to consider here. The first case occurs when $\r_i = L$.
In this case, since $U_i$ is a $\d{i}{o}$ closed cycle in $\K{L}^{i+2,i+1}$
and $\d{i}{o}$ is just the level $L$ Koszul-Tate differential, equation
$\FULLEXTENDEDKT$ implies that $U_i$ must be a $\d{i}{o}$ boundary.
Define
$$ Q_{i+1} = Y_i\eqn$$
keeping in mind that it is arbitrary up to a $\d{i}{o}$ boundary $U_i$. Since
$Y_i$ belongs to $\K{L}^{i+2,i+1}$ but only contains $c$'s (ghosts) of level
less than or equal to $L$, it must contain at least two  ghosts and an
antighost.  Similarly, the fact that the $b$-number is $i+1>L$ insures that
each monomial in $U_i$ contains at least two antighosts and one ghost.

The other case occurs when $\r_i = i$. In this case, we cannot define
$Q_{i+1} = Y_i$ as in the previous case if we are to satisfy equation
$\QBoundaryCond$ for $j=i+1$.  However, the inclusion $\K{i}\,\into\,\K{i+1}$
and equation $\YEQ$ implies that
$$ X_i = -\d{i+1}{o} Y_i\,.\  \tag\YEQTWO$$
Although $Y_i$ does not satisfy equation $\QBoundaryCond$, the previous
equation is invariant under the shift $Y_i\,\mapsto\,Y_i+U'_i$ where $U'_i$
is any $\d{i+1}{o}$ cocycle in $\K{i+1}^{i+2,i+1}$ so the question arises as
to whether $U'_i$ can be chosen so that $Y_i+U'_i$ satisfies the boundary
conditions.  Since the general form for $\d{i+1}{o}$ cocycles in
$\K{i+1}^{i+2,i+1}$ is given by equation $\GENCYCLES$ and we must satisfy
the boundary conditions in equation $\INDONE$, we conclude that
$$ Q_{i+1} = Y_i + \sum_{a_{i+1}=1}^{m_{i+1}} \c{i+1}{a_{i+1}}
(\sum_{a_i=1}^{m_i} \Z{i+1}{a_{i+1}}{a_{i}}\b{i}{a_i} +\M{i+1}{a_{i+1}}).\,\
\tag\NEXTQ$$
As in the previous case, the $Q_{i+1}$ is arbitrary up to $\d{i}{o}$
boundaries and each monomial in $Y_i$ contains at least two ghosts and one
antighost. Also, each monomial in $\M{i+1}{a_{i+1}}$ contains at
least two antighosts since $\M{i+1}{a_{i+1}}$ belongs to $K^{0,i+1}$ and
only contains antighosts $\b{j}{a_j}$ where $j\leq i-1$.  Therefore,
$Q_{i+1}$ is of the form
$$ Q_{i+1} = \sum_{a_{i+1}=1}^{m_{i+1}} \c{i+1}{a_{i+1}} (\sum_{a_i=1}^{m_i}
\Z{i+1}{a_{i+1}}{a_i}\b{i}{a_i} +\M{i+1}{a_{i+1}}) + {\rm etc}\ \eqn$$
where ${\rm etc}$ consists of terms with at least two antighosts and one
ghost or at least two ghosts and one antighost.

This takes care of the induction for equation $\INDONE$ but we still have to
perform the induction on $\INDTWO$. That is, we need to show that $R_{i+1}^2$
belongs to $F^{i+3}\K{i+1}$.  Let us begin by noting that the definition of
$X_i$ and $Q_{i+1}$ yields
$$ R_i^2 + \d{i+1}{o}Q_{i+1}\in F^{i+3}\K{i}^2\subseteq F^{i+3}\K{i+1}^2.\
\tag\RLEMMA$$
Since
$$\eqnalign{R_{i+1}^2 &= Q_{i+1}^2 + R_i^2 + \comm{R_i}{Q_{i+1}}\cr
                      &= - Q_{i+1}^2 + R_i^2 + \D{i+1} Q_{i+1} \cr
                      &= - Q_{i+1}^2 + R_i^2 + \d{i+1}{o}Q_{i+1} +
                           \d{i+1}{1}Q_{i+1} + \d{i+1}{2}Q_{i+1}+\cdots\
\tag\RIND}$$
and equation $\RLEMMA$ holds, we need only show that
$$ -Q_{i+1}^2 + \d{i+1}{1}Q_{i+1} + \d{i+1}{2}Q_{i+1}+\cdots\in\,
F^{i+3}\K{i+1}.\ \eqn$$
First of all, it is easy to see that $\d{i+1}{1}Q_{i+1} +
\d{i+1}{2}Q_{i+1}+\cdots$ belongs to $F^{i+3}\K{i+1}$ using the fact that
$Q_{i+1}$ belongs to $F^{i+2}\K{i+1}$ and the filtration degrees of
$\d{i+1}{j}$ for all $j > 0$.  Furthermore, $Q_{i+1}^2$ belongs to
$F^{i+3}\K{i+1}$ since the monomial in $Q_{i+1}^2$ with the lowest $c$-number
arises by taking a poisson bracket of a level $i$ ghost with a level $i$
antighost in computing $Q_{i+1}^2$ resulting in a term in $Q_{i+1}^2$ with
$c$-number $2(i+2) - (i+1) = i+3$.  (The reason that the commutator of a
level $i+1$ ghost with a level $i+1$ antighost does not appear in computing
$Q_{i+1}^2$ is that there are no level $i+1$ antighosts in $Q_{i+1}$.)
Therefore, we conclude that $Q_{i+1}$ belongs to $F^{i+3}\K{i+1}$.

This concludes the construction of the BRST charge.
\section{Classical BRST Cohomology}

In this section we will show that classical BRST cohomology vanishes for
negative ghost number and is isomorphic to the $E_2$ term in the spectral
sequence associated to the filtration by $c$-number by constructing the
explicit isomorphism extending the one given in ${\Ref\TKJMFReview}$.
Furthermore, at zero ghost number, we will show that classical BRST
cohomology is isomorphic as poisson algebras to $N(J)/J$ where $J$ is the
coisotropic ideal with respect to which the BRST complex was constructed. We
will discuss BRST cohomology at positive ghost number in the next section.

Let us begin by stating the basic result of this section.
\Thm{Let $\smooth{M}$ be a poisson algebra, $J$ be an associative ideal
generated by elements $\Psi=(\psi_1,\ldots,\psi_{m_o})$, and $ K$ be the
space of BRST cochains constructed relative to the constraints $\Psi$ and
$D: K^g\,\to\, K^{g+1}$ be the BRST differential given by the inner
derivation $D = \comm{Q}{\cdot}$ where $Q$ is the BRST charge.  Let $D$ be
decomposed by $c$-number
$$ D = \sum_{i=0}^\infty \dd_i\ \tag\DIFFDECOMP$$
where $\dd_i: K^{c,b}\,\to\, K^{c+i,b+i-1}$. If $H_D( K)$ is BRST
cohomology then there is an isomorphism of associative algebras
$$ H_D^g ( K) \cong \cases{{\bf 0} & for $g<0$\cr
                  H^g_{\dd_1}(H^0_{\dd_o}( K)) & for $g\geq 0$\cr}\,.\
\tag\ISO$$
If $g\geq 0$ and $\cl{x}$ is an element of $H_D^g$ then the isomorphism
$\chi: H_D^g( K)\,\isomap\,H^g_{\dd_1}(H^0_{\dd_o}( K))$ is given by
$$ \cl{x}\,\mapsto\,\cl{\cltwo{x_o}}\ \tag\ISOMAP$$
where $x_o$ is the component of $x$ in $ K^{g,0}$, $\cltwo{x_o}$ is an
element in $H_{\dd_o}$, and $\cl{\cltwo{x_o}}$ is an element in
$H^g_{\dd_1}(H^0_{\dd_o}( K))$.  In particular, at zero ghost number
$$ H_D^0( K) \cong {{N(J)}\over J}.\ \tag\SPECIALISO$$
The isomorphism in equation $\ISO$ can be used to define a poisson
superalgebra structure on $H^g_{\dd_1}(H^0{\dd_o}( K))$ which agrees with
the poisson algebra structure of $N(J)/J$ at zero ghost number.}

Before starting with the proof, note that $D^2 = 0$ is equivalent to the
string of equations
$$\sum_{i=0}^p \delta_i\delta_{p-i} = 0\ ,\tag\IDENTS$$
for each $p\geq 0$.  In particular, plugging in $p=0$ and $1$, we obtain
$$ \dd_o ^2 = 0\ \tag\IDENTZERO$$
and
$$ \dd_o\dd_1 + \dd_1\dd_o = 0.\tag\IDENTONE$$
We know that $\dd_o$ is just the Koszul-Tate differential from the
construction of the BRST charge in the previous section so equation
$\IDENTZERO$ is not too surprising.

Equation $\ISO$ arises from the fact that the BRST complex is a complex
filtered by $c$-number (see $\FILTRATION$).  This filtration has an
associated spectral sequence whose $E_o$ term is just the Koszul-Tate complex
so that the $E_1$ term is just $H_{\dd_o}( K)$.  However, the $E_2$ term in
the spectral sequence is the cohomology of the complex $H_{\dd_o}(K)$
with the differential induced by $\dd_1$ which we shall also denote by
$\dd_1$. Therefore, $E_2^{c,b}$ is just $H^c_{\dd_1}(H^b_{\dd_o}( K))$.
The spectral sequence degenerates at this point because of the acyclicity of
the Koszul-Tate complex.  Notice that if $b>0$ then $E_2^{c,b}$ vanishes
because of the acyclicity of the Koszul-Tate complex.  If the constraints
$\Psi$ are irreducible, then the filtration is bounded since the BRST complex
is finite dimensional.  In this case, we know that the $E_2$ term is
isomorphic to $H_D$ and the fact that $E_2^{c,b} = 0$ for $b>0$ implies that
$H_D^g = 0$ for all $g<0$.  However, in the case where $\Psi$ are reducible
constraints, this filtration is no longer bounded and, therefore, it is not
immediately clear if the $E_2$ term is isomorphic to $H_D$.   We will first
show that $H_D$ vanishes for negative ghost number directly and then show
that the map $\ISOMAP$ between $H_D$ and the $E_2$ term is, in fact, an
isomorphism.

Let us now assume that $x$ is a BRST cocycle in $ K^g$
where $g < 0$.  We can decompose $x$ by $c$-number to get $x =
\sum_{i=0}^\infty x_i$ where $x_i$ belongs to $ K^{i,i-g}$. Decomposing the
equation $Dx = 0$ by $c$-number yields
$$ \sum_{j=0}^i\dd_j x_{i-j} = 0\ \tag\DCLOSED$$
for all $i\geq 0$.  Plugging in $i=0$ and $1$, for example, yields
$$ \dd_o\, x_o = 0\,,\tag\XONE$$
and
$$ \dd_1\, x_o + \dd_o\, x_1 = 0\,.\tag\XTWO$$
We will show that there exists a $y$ in $ K^{1-g}$ such that $x = Dy$.
We decompose $y$ by $c$-number to get $y = \sum_{i=0}^\infty y_i$ where
$y_i$ belongs to $ K^{i,1-g+i}$ for all $i\geq 0$.  Decomposing the equation
$x=Dy$ by $c$-number tells us that such a $y$ exists if and only if there
exist $y_i$ in $ K^{i,1-g+i}$ which satisfy
$$ x_p = \sum_{j=0}^p\dd_j y_{p-j}\ \tag\DEXACT$$
for all $p \geq 0$.

The existence of such $y_i$ is a consequence of the acyclicity of the
Koszul-Tate complex.  For example, equation $\XONE$ tells us that $x_o$ is a
$\dd_o$ closed cycle and, therefore, $\dd_o$ exact from the acyclicity of the
Koszul-Tate complex since $x_o$ has antighost number of at least one.  In
other words, there exists $y_o$ in $K^{0,1-g}$ such that $ x_o = \dd_o y_o$
which is just equation $\DEXACT$ where $p=0$.  Similarly, equations $\XTWO$
and $\IDENTONE$ implies that
$$ 0 = \dd_o x_1 + \dd_1 x_o = \dd_o x_1 + \dd_1 \dd_o y_o = \dd_o (x_1 -
\dd_1 y_o).\ \eqn$$
The acyclicity of the Koszul-Tate complex tell us that there exists $y_1$ in
$K^{1,2-g}$ such that $ x_1 - \dd_1 y_o = \dd_o y_1$ which is just equation
$\DEXACT$ with $p = 1$.

The construction of the higher terms in $y$ proceeds by induction.  Suppose
that there exists $y_o, y_1, \ldots, y_i$ satisfying equation $\DEXACT$ for
all $p = 0,\ldots,i$ then
$$\eqnalign{ &\dd_o (x_{i+1} - \sum_{j=1}^{i+1} \dd_j y_{i+1-j}) =\cr
& = \dd_o x_{i+1} + \sum_{j=1}^{i+1}\sum_{l=1}^l\dd_l\dd_{j-l}
    y_{i+1-j}&\hbox{using $\IDENTS$}\cr
& = \dd_o x_{i+1} + \sum_{l=1}^{i+1}\sum_{j=l}^{i+1}\dd_l\dd_{j-l}
    y_{i+1-j}\cr
& = \dd_o x_{i+1} + \sum_{l=1}^{i+1}\dd_l\sum_{s=0}^{i+1-l}\dd_s
    y_{i+1-s-l}\cr
& = \dd_o x_{i+1} + \sum_{l=1}^{i+1}\dd_l x_{i+1-l}&\hbox{from induction
    hypothesis}\cr
& = \sum_{l=0}^{i+1}\dd_l x_{i+1-l} = 0\ \tag\KEYSTEP}$$
where we have used that fact that $x$ is a $D$ cocycle in the last step.
Since $x_{i+1}-\sum_{j=1}^{i+1} y_{i+1-j}$ is a $\dd_o$ closed cycle with
antighost number $i+1-g>0$, the acyclicity of the Koszul-Tate complex tells
us that there exists a $y_{i+1}$ in $K^{i+1,2-g+i}$ such that
$$ x_{i+1} = \sum_{j=1}^{i+1} \dd_j y_{i+1-j} +  \dd_o y_{i+1} =
\sum_{j=0}^{i+1} \dd_j y_{i+1-j}.\ \eqn$$
This completes the induction.  Therefore, BRST cohomology vanishes for
negative ghost number.

Let us now assume that $g\geq 0$.  We will show that the map $\ISOMAP$ is
an isomorphism of associative algebras.  We first check that the map is well
defined.  Consider any $y\in K^g$ for $g\ge 0$ then we have the decomposition
$y=y_0+y_1+\cdots$ where $y_i\in K^{g+i,i}$ for all $i\ge 0$.  In this case,
$Dy$ belongs to $ K^{g+1}$ and the component of $Dy$ in $ K^{g+1,0}$ is given
by $\delta_1\, y_0+\delta_0\, y_1$. We see that
$\chi(\cl{Dy}) = \cltwo{\clthree{\delta_1\,y_0+ \delta_0\, y_1}} =
\cltwo{\clthree{\dd_1\, y_0}} = 0$ and, therefore, the map is well-defined.

Is $\chi$ injective?  Consider $x\in K^g$ which decomposes
into $x=x_0+x_1+x_2+\ldots$ where $x_i\in K^{g+i,i}$ such that
$\chi(\cl{x})=\cltwo{\clthree{x_0}}=0$.  We need to show that there exists
$y\in K^{g-1}$ such that $x=D y$.  Decomposition of the previous equation is
equivalent the existence of $y_i\in K^{g-1+i,i}$ such that for all $p\ge 0$,
$$x_p=\sum_{i=0}^{p+1} \delta_i\, y_{p+1-i}\,.\ \tag\DClosedEquation$$
The fact that $\cltwo{\clthree{x_0}}=0$ implies that
$\clthree{x_0}=\dd_1\,\clthree{y_0}$ with some $y_0\in K^{g-1,0}\,$ or
$\clthree{x_0-\delta_1\, y_0}=0$. Therefore, $x_0-\delta_1\,
y_0-\delta_0\, y_1 = 0$ for some $y_1\in K^{g,1}$.  In other words,
$$ x_0 = \delta_0\,y_1 + \delta_1\,y_0\ \eqn$$
for some $y_0$ and $y_1$.  This is just \equ $\DClosedEquation$
for $p=0$.  We now proceed to show $\DClosedEquation$ by induction.  Suppose
that there exist $y_i\in K^{g-1+i,i}$ for all $i=0,\ldots,r$ which satisfy
$\DClosedEquation$ for all $p=0,\ldots,r-1$ then a similar argument to
equation $\KEYSTEP$ yields
$$\delta_0\,(x_r-\sum_{j=1}^{r+1}\delta_j\,y_{r+1-j}) = 0.\ \eqn$$
Therefore, there exists an element $y_{r+1}\in K^{g+r,r+1}$ which satisfies
$$x_r=\sum_{j=0}^{r+1}\delta_j\,y_{r+1-j}\, .\eqn$$
This completes our induction. The proof of the surjectivity of $\chi$
proceeds similarly.

Finally, it is clear that $\chi$ is an isomorphism of associative
algebras since if $x\in K^g$ and $y\in K^h$ which decompose into sums
of $x_i\in K^{g+i,i}$ and $y_i\in K^{h+i,i}$, respectively, then
$\chi(\cl{x}\cl{y}) = \chi(\cl{x y}) = \cltwo{\clthree{x_0 y_0}}
  = \chi(\cl{x})\chi(\cl{y})\, .$
We can endow $H_{\dd_1}(H_{\dd_o}(K))$ with the structure of a poisson
superalgebra by defining the poisson bracket on elements $x,y\in
H_{\dd_1}(H_{\dd_o}(K))$  by
$$\comm{x}{y} = \chi(\comm{\chi^{-1}(x)}{\chi^{-1}(y)}).\ \eqn$$

\section{Vertical Cohomology Is Not BRST Cohomology}

In this section, we show that in the case of the reduction of a poisson
manifold by a submanifold,  BRST cohomology is not generally isomorphic
to the cohomology of vertical differential forms with respect to the null
foliation. We prove this by looking at a  simple counterexample.

Let $N$ be a smooth manifold and $\F\,\to\,N$ be a smooth involutive rank $k$
subbundle of the tangent bundle $TN\,\to\,N$.  Let us denote the space of
leaves of the associated foliation by $\Ntilde$ and the canonical projection
map, which need not be smooth, by $\pi:N\,\to\,\Ntilde$.  The vectors in
$\F_n$ for all points $n$ in $N$ are called the {\sl vertical vectors at $n$
with respect to the foliation $\pi:N\,\to\,\Ntilde$,} \eg\  if $\pi$ is a
smooth map then $\F_n$ are the vectors in $T_n N$ in the kernel of $\pi_*$.
The space of {\sl vertical differential forms with respect to the foliation
$\pi:N\,\to\,\Ntilde$}, $\Omega_\F(N)$, is defined to be sections of the
bundle $\exterior\F\,\to\,N$.  There exists a natural map $\Omega^p(N)\,\to\,
\Omega^p_\F(N)$, denoted by $\gamma\,\mapsto\,\gammabar$, by restricting
$\gamma$ to act upon only vectors in $\F$.  This map is surjective.  A
consequence is that the exterior derivative $d:\Omega^p(N)\,\to\,
\Omega^{p+1}(N)$ induces a derivative operator (called the {\sl vertical
derivative}) $d_\F:\Omega_\F^p(N)\,\to\,\, \Omega_\F^p(N)$ via $d_\F\,
\gammabar = \overline{d\gamma}$.  The vertical derivative $d_\F$  is
well-defined if $\F$ is involutive.$^{\Ref\Warner}$  This is the case because
the kernel of the above map $\Omega^p(N)\,\to\, \Omega^p_\F(N)$ forms a
differential ideal in $\Omega(N)$.  The cohomology of this complex $H_\F(N)$
is called the {\sl cohomology of vertical differential forms with respect to
the foliation $\pi:N\,\to\, \Ntilde$.}

Let $(\Mtilde,\Ptilde)$ be the reduction of the poisson manifold $(M,P)$ by
the submanifold $M_o$ then the space of vertical vectors at $m$ in $M_o$ with
respect to foliation $M_o\,\to\,\Mtilde$ is just $V_m = T_m M_o\intersect T_m
M_o^\perp$.  Clearly, $V\,\to\,M_o$ forms an involutive subbundle of $TM_o$
and we have $d_V:\Omega_V^p(M_o) \,\to\,\Omega_V^{p+1}(M_o)$, the complex of
vertical differential forms with respect to the foliation
$\pi:M_o\,\to\,\Mtilde$. (Notice that $\pi$ need not be smooth here.) More
explicitly, the vertical derivative is given by
$$ (d_V f)(X) = X(f),\ \eqn$$
$$ (d_V \alpha)(X,Y) = X(\alpha(Y)) - Y(\alpha(X)) - \alpha(\comm{X}{Y}),\
\eqn$$
and
$$ d_V (\omega\wedge\psi) = (d_V\omega)\wedge\psi + (-1)^{\abs{\omega}} \omega
\wedge(d_V\psi)\ \eqn$$
for all $f$ in $\smooth{M_o}$, $X,Y$ in $\sec{TM_o^\perp}$, $\alpha$ in
$\Omega_V^1(M_o)$, and $\omega$, $\psi$ in $\Omega_V(M_o)$ with definite
degree.  The cohomology of this complex is denoted by $H_V(M_o)$.

Let us now consider the special case where $(M,P)$ is a symplectic manifold
and $I$ is the ideal of functions in $\smooth{M}$ which vanish on $M_o$.
Recall that $\sec{V} = \sec{TM_o^\perp\intersect TM_o}$ consists of the
restriction of hamiltonian vector fields of elements in $I'$ to $M_o$.
In other words, there is a surjective homomorphism of lie algebras $I'
\,\to\, \sec{V}$ given by $i\,\mapsto\,X_i|_{M_o}$.  Furthermore, this map
has a kernel since $ 0 = X_i|_{M_o} = \Psh di|_{M_o} $ but
$\Psh:\Ann(M_o)\,\to\,TM_o^\perp$ is invertible since $(M,P)$ is symplectic.
Therefore, $di|_{M_o} = 0$ but $di|_{M_o}$ is a section of the annihilator
bundle of $M_o$.  The isomorphism in equation $\ANNISO$ implies that $i$ must
be an element of $I^2$.  In other words, we have the isomorphism of lie
algebras
$$ {{I'}\over {I^2}} \isomap \sec{V}\ \eqn$$
which is also a $\smooth{M}/I\,\cong\,\smooth{M_o}$ module isomorphism.
Therefore, we can make the identification
$$ \Omega_V(M_o) \,\cong\,\Evert{I}^*\ \eqn$$
where $\Evertp{p}{I}^*$ consists of all alternating $\smooth{M}/I$-linear
$p$-forms from $I$ to\nl\message{unnatural page break}
 $\smooth{M}/I$.  The vertical differential in this setting can be identified
with
$$ (\delta_V \cl{f})(\cltwo{\phi_1}) = \cl{\comm{f}{\phi_1}},\ \tag\VDIFFONE$$
$$ (\delta_V\beta)(\cltwo{\phi_1},\cltwo{\phi_2}) =
\cl{\comm{\phi_1}{\beta(\cltwo{\phi_2})}} -
\cl{\comm{\phi_2}{\beta(\cltwo{\phi_1})}} -
\cl{\beta(\cltwo{\comm{\phi_1}{\phi_2}})},\ \tag\VDIFFTWO$$
for all $\beta$ in $(I'/I^2)^*$, $\cl{f}$ in $\smooth{M}/I$ and
\message{2/23/92 -- NEW BRACKETS IN BETA TO FIX TYPO!}
$\cltwo{\phi_1}, \cltwo{\phi_2}$ in $I'/I^2$ and then extending it via
derivation to $\exterior_{\smooth{M}/I}(I'/I^2)^*$.  This algebraic
formulation of vertical cohomology can also be extended to the case of the
reduction of poisson manifolds by defining $\C(I) = \{i\in
I'\,|\,\comm{i}{g}\in I,\  \forall\,g\in\smooth{M}\}$ and then replacing
$I^2$ by $\C(I)$ in the above since $\C(I)$ contains all of the elements $i$
in $I'$ such that $X_i|_{M_o} = 0$.  Such a complex is well-defined for any
ideal $J$ in $\smooth{M}$.  Let us call this algebraic complex
$\delta_V:\exterior^p_{\smooth{M}/{\C(J)}}(J'/{\C(J)})^*\,\to
\exterior^{p+1}_{\smooth{M}/{\C(J)}}(J'/{\C(J)})^*$ the {\sl vertical
complex of $\smooth{M}$ with respect to $J$}.  Let us call the cohomology of
this complex $H_{\delta_V}$ the {\sl vertical cohomology of $\smooth{M}$ with
respect to $J$}.

In the very special case where $(\Mtilde,\Ptilde)$ is the symplectic
reduction of the symplectic manifold $(M,P)$ by the coisotropic submanifold
$M_o$ then we have $I' = I$ and $\C(I) = I^2$.  In other words, the cochains
in vertical cohomology are given by $\Evertical{I}^*$ and $\delta_V$ is still
given by the formulas above.  This gives rise to yet another algebraic
cohomology theory.  Let  $J$ be a coisotropic ideal in $\smooth{M}$ then we
can define the complex $\db:\Everticalp{p}{J}\,\to\,\Everticalp{p+1}{J}$
where the differential $\db$ is defined by equations $\VDIFFONE$ and
$\VDIFFTWO$ (where $\delta_V$ is replaced by $\db$) and then extended as a
graded derivation.  Let us denote the cohomology of this complex by
$H_{\db}$.  $H_{\db}$ is an example of {\sl Rinehart
cohomology}.$^{\Ref\Rinehart}$

How does all this relate to BRST cohomology, $H_D$?  We showed in the
previous section that $H_D$ is isomorphic to $H_{\dd_1}(H_{\dd_o})$ which is
isomorphic to $H_V(M_o)$ in the case of the reduction of a symplectic
manifold $(M,P)$ by a closed and embedded coisotropic submanifold $M_o$
provided that the normal bundle of $M_o$ is trivial.$^{{\Ref\DuVi},
{\Ref\HT}, {\Ref\TKJMFReview}}$  The results of ${\Ref\FHST}$ and
${\Ref\Stashefftwo}$ suggests that this is the case even if the normal bundle
is nontrivial.  However, it is not true that for the case of reduction of a
general poisson manifold, BRST cohomology can be identified with vertical
cohomology.  Consider the example where $(M,P)$ is a poisson  manifold with
the trivial poisson structure $P=0$ and $M_o$ is a closed and embedded
submanifold of $M$.  In this case, $\sec{TM_o^\perp} = 0$ by definition so
that there are no vertical vectors so that $\Omega_V(M_o)=\smooth{M_o}$. Let
$J=I$ be the ideal of functions in $\smooth{M}$ which vanish on $M_o$. Let us
furthermore assume that $I$ is an ideal generated by a collection of
irreducible constraints $\Psi=(\psi_1,\ldots, \psi_{m_o})$. In this case, the
BRST operator is given by
$$ Q = \sum_{i=1}^{m_o} \c{o}{i}\psi_i\ \eqn$$
then the BRST differential is given by
$$ D = \sum_{i=1}^{m_o} \psi_i\comm{\c{o}{i}}{\cdot}\ \eqn$$
which is precisely the Koszul differential, \ie\  $D = \dd_o$ and $\dd_i=0$
for all $i>0$.  Therefore, BRST cohomology is given by
$$ H_D^g \cong \cases{{\bf 0} & for $g< 0$\cr
               \Symmetric^g_{\smooth{M}/J}(V_1)^* &for $g\geq 0$\cr}\,\eqn$$
where $V_1$ is the $m_o$-dimensional free $\smooth{M}/I$ module of level zero
antighosts and \nl\message{unnatural page break}
$\Symmetric^g_{\smooth{M}/J}(V_1)^*$ is the space  of ghost number $g$
elements of the symmetric superalgebra over $V_1$. Therefore, it is not
correct to identify vertical cohomology with BRST cohomology here. However,
it would still be correct to make the identification
$$ H^g_D \cong H^g_{\db}\ \tag\BRSTPOISSON$$
if $g\geq 0$ since if $J$ is generated by irreducible elements $\Psi$ then by
equation $\IRREDUCIBLE$, $J/J^2$ is a free $m_o$-dimensional
module over $\smooth{M}/J$ so that
$\Evertical{J}^*\isomap\Symmetric_{\smooth{M}/J}(V_1)^*$.  Of course, this
isomorphism is true since  $\Symmetric$ is the symmetric superalgebra which
is, in this case, an exterior (nonsuper)algebra since elements in $V_1$ have
antighost number $1$.

The question then arises as to whether this correspondence holds in general.
That is, if $H_D$ is the  BRST cohomology associated to the reduction of the
poisson algebra $\smooth{M}$ by a coisotropic ideal $J$ then is it true
the isomorphism in equation $\BRSTPOISSON$ holds?
This correspondence is certain true in the case of symplectic reduction of
$\smooth{M}$ by a coisotropic ideal $J=I$ of functions which vanish on a
coisotropic submanifold as well as for the simple example given above.  It
remains to be seen whether it is true in general.

\section{Conclusion}

In this paper, we have generalized classical BRST cohomology to the more
general framework of the reduction of a poisson algebra $\smooth{M}$ by a
coisotropic ideal.  This setting encompasses the reduction of poisson
manifolds.  Let us make a few remarks.

First of all, it is not known what classical BRST cohomology computes for
positive ghost numbers, in general.  Is it isomorphic to this cohomology
theory of Rinehart?

Secondly, most of the constructions in this paper extend to the case where
${\cal P}$ is a poisson algebra which is a Noetherian ring under associative
multiplication, \ie\ when all ideals of $\P$ are finitely generated. This is
not the case, for example, when $\P$ is the poisson algebra of smooth
functions on an infinite-dimensional poisson manifold. An extension of
classical BRST cohomology to this case would be enlightening especially in
applications to classical field theory.

Thirdly, BRST cohomology has a version which appears in quantum theory,
usually in the context of lie algebra cohomology.$^{{\Ref\KS}, {\Ref\FGZ},
{\Ref\FF}}$ There has been much work relating quantum BRST cohomology to
classical BRST using methods of quantization inspired by geometric
quantization,  \eg\ ${\Ref\TKSeattle}, {\Ref\TKJMFGQ}, {\Ref\Hu},
{\Ref\DEGST}, {\Ref\Tuynman}, {\Ref\DET}$. An extension of the results in
this paper to the case of geometric quantization is in
progress.$^{\Ref\TKPreview}$.

Finally, an extension of these techniques to the case of the reduction of a
poisson supermanifold would be useful in certain physical applications, \eg\
the covariant quantization of the superstring.

\refsout
\bye